\documentstyle[preprint,aps,epsfig]{revtex}
\tighten
\begin{document}
\draft
\preprint{\vbox{Submitted to Physical Review C \hfill FSU-SCRI-95-105 \\
                                          \null\hfill nucl-th/9510024}}
\title{Stability Analysis of the Instantaneous Bethe-Salpeter Equation and
the Consequences for Meson Spectroscopy}
\author{J.~Parramore}
\address{Department of Physics and
         Supercomputer Computations Research Institute, \\
         Florida State University, Tallahassee, FL 32306}
\author{H.-C. Jean}
\address{Department of Physics and
         Supercomputer Computations Research Institute, \\
         Florida State University, Tallahassee, FL 32306}
\author{J.~Piekarewicz}
\address{Supercomputer Computations Research Institute, \\
         Florida State University, Tallahassee, FL 32306}
\date{\today}
\maketitle
\begin{abstract}
We investigate the light and heavy meson spectra in the context of
the instantaneous approximation to the Bethe-Salpeter equation
(Salpeter's equation). We use a static kernel consisting of a
one-gluon-exchange component and a confining contribution.
Salpeter's equation is known to be formally equivalent to a
random-phase-approximation equation; as such, it can develop
imaginary eigenvalues. Thus, our study can not be complete
without first discussing the stability of Salpeter's equation.
The stability analysis limits the form of the kernel and reveals
that, contrary to the usual assumption, the confining component
can not transform as a Lorentz scalar; it must transform as the
timelike component of a vector. Moreover, the stability analysis
sets an upper limit on the size of the one-gluon-exchange component;
the value for the critical coupling is determined through a solution
of the ``semirelativistic'' Coulomb problem. These limits place
important constraints on the interaction and suggest that a more
sophisticated model is needed to describe the light and heavy quarkonia.
\end{abstract}
\pacs{PACS number(s):~11.10.St,12.40.Qq,14.40.-n}
\section{Introduction}
In hadron-structure theory one is interested in describing the hadron
as a relativistic composite system. To date, most basic properties of
hadrons cannot yet be derived from QCD---the fundamental theory of
the strong interactions. Note, however, that QCD sum rules can place
some constraints regarding quark-distribution amplitudes in mesons and
baryons~\cite{shifman79,reinders85}. With the advent of more powerful
computing facilities, lattice gauge theory~\cite{rothe92} should
provide an increasingly useful means of studying hadronic physics.
Yet, at the present time it does not provide a convenient framework
for a systematic study of a large variety of hadronic phenomena.
Specifically, with the commission of state-of-the-art facilities,
such as CEBAF\footnote{The Continuous Electron Beam Accelerator
Facility}, other nonperturbative techniques will be required which
can be used to incorporate phenomena at many different length scales
within a single theoretical framework.

To a large extent our current understanding of hadronic structure is
based on the nonrelativistic constituent quark
model~\cite{gellmann64,godfrey89}. Meson properties are well reproduced
by a phenomenological potential consisting of the sum of a short-range
one-gluon exchange (OGE) component and a long-range confining
contribution. A quantitative description of meson masses, their static
properties and decay rates count among the many successes of the model.
Generally, one would prefer to have a relativistic and manifestly covariant
model. For example, a covariant formalism will enable one to relate
the wave function (or vertex function) in different frames. This
becomes essential for calculating hadronic form factors at finite
momentum transfer.

The starting point for most relativistic studies of the meson
spectrum is the covariant Bethe-Salpeter equation~\cite{salpeter51}.
The Bethe-Salpeter equation can be regarded as the relativistic
generalization of the Lippmann-Schwinger equation. However, the
Bethe-Salpeter equation, being covariant, depends on the zeroth
component of the relative four-momentum (i.e., the relative energy).
Aside from the technical difficulties encountered in handling this
extra degree of freedom, one must decide in the present case how to
generalize the essentially nonrelativistic quark-antiquark potential
to four dimensions---a nontrivial task to carry out correctly. The
difficulty in dealing with the relative energy has led to many different
approximations to the Bethe-Salpeter equation wherein one works within
a three-dimensional reduction but attempts to retain fundamental
physical principles. There is no obviously correct method. Thus, one
should study different three-dimensional reductions in the hope of
isolating model-independent results. Here we work within the
instantaneous Bethe-Salpeter framework (Salpeter's framework). Although
retardation effects and manifest covariance are lost, one retains
relativistic kinematics, the relativistic character of the potential,
and the Dirac structure of positive- and negative-energy states.

The use of the instantaneous approximation commonly employed in the
literature entails other problems besides the loss of retardation
and manifest covariance;  Salpeter's bound-state equation is represented
by a {\it nonhermitian} Hamiltonian. Indeed, it has been
recognized~\cite{long84,resag93,parra94,piek92} that Salpeter's equation
is identical in structure to a random-phase-approximation (RPA) equation
familiar from the study of nuclear collective excitations~\cite{thou60}.
Thus, it can be rewritten as a hermitian eigenvalue equation---but
for the square of the energy. This suggests the possibility of
imaginary eigenvalues which would signal the onset of an instability.
For our problem of interest these imaginary solutions are unphysical
and their appearance can be precluded by limiting the form of the
kernel. This is achieved through a stability analysis of Salpeter's
equation. We should stress that any study based on Salpeter's equation
is not complete until the stability analysis is performed. The main
goal of this paper is to present the stability analysis and to examine
the implications for the meson spectra.

  We have organized the paper as follows. In Sec.~\ref{secformal},
Salpeter's equation is presented and the method used to solve it
is reviewed.  In Section~\ref{secstability}, we study the interaction
kernel for the particular Lorentz structures of interest. The
stability analysis for the confining part of the kernel is
reviewed~\cite{parra94} and the analysis for the OGE component
developed. As a result of the stability analysis the form of the
instantaneous kernel is constrained. We examine the consequences
of these constraints on the heavy- and light-meson spectroscopy
in Section~\ref{secspectra}. Finally, our concluding remarks
are presented in Section~\ref{secconcl}.
\section{Formalism}
\label{secformal}
\subsection{Salpeter's Equation}
In the Salpeter formalism~\cite{salpeter52}, the bound-state
spectrum is generated as a solution to the instantaneous Bethe-Salpeter
equation in the ladder approximation.  In this approximation, the
irreducible Bethe-Salpeter kernel is given by
\begin{equation}
  V(x_{1},x_{2}) \equiv V({\bf x_{1}},{\bf x_{2}}) \;.
 \label{vinst}
\end{equation}
In Ref.~\cite{parra94}, the derivation of Salpeter's equation was
illustrated using Green's function methods, which yielded the
following eigenvalue equation for the Salpeter
wave function $\chi^{E}$:
\begin{equation}
  \chi^{E}_{\alpha\sigma}({\bf x}_1,{\bf y}_2) =
   \int d^3 z_1 d^3 z_2 \;
   G^{(0)}_{\alpha\eta^{\prime};\xi\sigma}
   ({\bf x}_1,{\bf z}_2;{\bf z}_1,{\bf y}_2;E)
   V_{\xi\eta ;\xi^{\prime}\eta^{\prime}}({\bf z}_1,{\bf z}_2)
   \chi^{E}_{\xi^{\prime}\eta}({\bf z}_1,{\bf z}_2) \;\; ,
\label{salpchi}
\end{equation}
where the Salpeter wave function is defined by
\begin{equation}
  \chi^{E}_{\alpha\sigma}({\bf x}_1,{\bf y}_2) \equiv
  <{\Psi_{0}}|\psi_{\alpha}({\bf x}_{1}) \bar{\psi}_{\sigma}({\bf y}_{2})
  |{\Psi_{E}}> \;\; ,
\end{equation}
$G^{(0)}$ is the free two-body propagator in the instantaneous approximation
(IA), $\Psi_{0}$ represents the vacuum, and $\Psi_{E}$ represents the bound
state with energy $E$. Expanding the fermion fields in a single-particle
basis and using the properties of the free two-body Green's function
then gives
\begin{eqnarray}
   \chi^{E}_{\alpha\sigma}({\bf x}_1,{\bf y}_2)
 = \sum_{{\bf k}_{1}s_{1};{\bf k}_{2}s_{2}}
   \bigg(
     &&
     \Big[U_{{\bf k}_{1}s_{1}}({\bf x}_1)\Big]_\alpha
     \Big[\bar{V}_{{\bf k}_{2}s_{2}}({\bf y}_2)\Big]_\sigma
      B_{s_{1}s_{2}}({\bf k}_{1},{\bf k}_{2}) + \nonumber \\
     &&
     \Big[V_{{\bf k}_{1}s_{1}}({\bf x}_1)\Big]_\alpha
     \Big[\bar{U}_{{\bf k}_{2}s_{2}}({\bf y}_2)\Big]_\sigma
      D_{s_{1}s_{2}}({\bf k}_{1},{\bf k}_{2})
   \bigg) \;,
\label{chidef3}
\end{eqnarray}
where the Salpeter amplitudes $B$ and $D$ are defined by
\begin{eqnarray}
   B_{s_{1}s_{2}}({\bf k}_{1},{\bf k}_{2}) &\equiv&
     \langle \Psi_{0} |
       b_{{s}_{1}}({\bf k}_{1})
       d_{{s}_{2}}({\bf k}_{2})
     | \Psi_{E} \rangle \;, \\
   D_{s_{1}s_{2}}({\bf k}_{1},{\bf k}_{2}) &\equiv&
     \langle \Psi_{0} |
       d^{\dagger}_{{s}_{1}}({\bf k}_{1})
       b^{\dagger}_{{s}_{2}}({\bf k}_{2})
     | \Psi_{E} \rangle \;,
\end{eqnarray}
and contain all dynamical information about the bound state.

Salpeter's equations are more conveniently expressed in an angular momentum
basis.  Projecting out the Salpeter amplitudes, expressing Salpeter's
equations for $B$ and $D$ in the center of momentum frame, and introducing the
partial-wave decomposition of the amplitudes in terms of total $L$ and $S$
coupled to the total angular momentum $J$ of the bound state
\begin{eqnarray}
\lefteqn{B_{s_1 s_2}({\bf k})} \nonumber \\
&&= \sum_{ S M_S L M_L J M}
      \langle {\frac{1}{2} s_1 ; \frac{1}{2} s_2}\vert{S M_S}\rangle
      \langle {L M_L ; S M_S} \vert {J M} \rangle
      Y_{L,M_L}(\hat{k}) B_{LSJM}(k) \;\; ,
\label{bvecwave}
\end{eqnarray}
\begin{eqnarray}
\lefteqn{(-)^{1 - s_1 - s_2} D_{-s_1 -s_2}({\bf k})}  \nonumber \\
&&= \sum_{S M_S L M_L J M}
      \langle {\frac{1}{2} s_1 ; \frac{1}{2} s_2}\vert{S M_S}\rangle
      \langle {L M_L ; S M_S} \vert {J M} \rangle
      Y_{L,M_L}(\hat{k}) D_{LSJM}(k) \;\; ,
\label{dvecwave}
\end{eqnarray}
one can write Salpeter's equations in an angular momentum basis as
\begin{eqnarray}
  \left[ +E - 2E_{k} \right] b_{LSJ}(k)
   =
  \int_{0}^{\infty} \frac{d{k^\prime}}{(2\pi)^3}
  \sum_{L^\prime S^\prime}
   & \biggl \{ &
       \langle{k;LSJ}
       \vert {V}^{++} \vert {k^\prime;L^\prime S^\prime J} \rangle
                   b_{L^\prime S^\prime J}(k^\prime) \nonumber \\
     &+& \langle{k;LSJ}
       \vert {V}^{+-} \vert {k^\prime;L^\prime S^\prime J} \rangle
                   d_{L^\prime S^\prime J}(k^\prime)
     \biggl \} \;,
\label{bklsj}
\end{eqnarray}
\begin{eqnarray}
  \left[ -E - 2E_{k} \right] d_{LSJ}(k)
   =
  \int_{0}^{\infty} \frac{d{k^\prime}}{(2\pi)^3}
  \sum_{L^\prime S^\prime}
   & \biggl \{ &
       \langle{k;LSJ}
       \vert {V}^{-+} \vert {k^\prime;L^\prime S^\prime J} \rangle
                   b_{L^\prime S^\prime J}(k^\prime)   \nonumber \\
   &+& \langle{k;LSJ}
       \vert {V}^{--} \vert {k^\prime;L^\prime S^\prime J} \rangle
                   d_{L^\prime S^\prime J}(k^\prime)
     \biggl \} \;,
\label{dklsj}
\end{eqnarray}
with $b(k) \equiv kB(k)$ and $d(k) \equiv kD(k)$.  For local interactions,
such as the ones considered here, the matrix elements of the potential are
given by (a sum over greek indices is implicitly assumed, and
$\bar{\alpha} \equiv 1-\alpha$)
\begin{eqnarray}
   \langle{k;LSJ}
   \vert {V}^{++} \vert {k^\prime;L^\prime S^\prime J} \rangle
 = \langle{k;LSJ}
   \vert {V}^{--} \vert {k^\prime;L^\prime S^\prime J} \rangle
   \nonumber \\
 = \sum_{{\cal L}{\cal S}}
   {(-1)}^{\alpha + \beta} {\cal F}_{{\cal L}{\cal S};LSJ}^{\alpha\beta}(k)
   \langle {{\cal{S}}} \vert\vert
   {\left[V_{\cal{L}}(k,k^\prime)
     \right]}_{\alpha\beta ;{\alpha^\prime}{\beta^\prime}}
   \vert\vert {{\cal{S}}} \rangle
   {\cal F}_{{\cal L}{\cal S};L^\prime S^\prime J}^
   {\alpha^\prime \beta^\prime}(k^\prime) \;,
\end{eqnarray}
\begin{eqnarray}
   \langle{k;LSJ}
   \vert {V}^{+-} \vert {k^\prime;L^\prime S^\prime J} \rangle
 = \langle{k;LSJ}
   \vert {V}^{-+} \vert {k^\prime;L^\prime S^\prime J} \rangle
   \nonumber \\
 = \sum_{{\cal L}{\cal S}}
   {(-1)}^{\alpha + \beta + {\cal L}}
   {\cal F}_{{\cal L}{\cal S};LSJ}^{\alpha\beta}(k)
   \langle {{\cal{S}}} \vert \vert
   {\left[V_{\cal{L}}(k,k^\prime)
     \right]}_{\alpha\beta ;{\alpha^\prime}{\beta^\prime}}
   \vert\vert {{\cal{S}}} \rangle
   {\cal F}_{{\cal L}{\cal S};L^\prime S^\prime J}^
   {\bar{\alpha}^\prime \bar{\beta}^\prime}
          (k^\prime) \;.
\end{eqnarray}
where
\begin{equation}
   {\cal F}_{{\cal L}{\cal S};L S J}^{\alpha \beta}(k)
   = C_{\alpha\beta}(k)
     \sum_{\lambda}
     \langle{\alpha 0 ; \beta 0}\vert{\lambda 0}\rangle
     \langle{{\cal L}{\cal S}J}\vert\vert
     {\left
     [Y_{\lambda}
     {\left(\sigma_\alpha \sigma_\beta \right)}_{\lambda}\right]}_0
     \vert\vert{LSJ}\rangle \;\; ,
\end{equation}
\begin{equation}
  C_{\alpha\beta}(k)
  = \sqrt{4\pi} {(-1)}^\alpha
    {\left[ \frac{E_{k} + M}{2E_{k}} \right]}
    \xi_{\alpha}(k) \xi_{\beta}(k) \; ; \;
    \xi_{\alpha}(k) = \left \{ \begin{array}{ll}
                          1 & \mbox{if $\alpha$ = 0 ;} \\
                     \frac{k}{E_{k} + M} & \mbox{if $\alpha$ = 1.}
                               \end{array}
                      \right.
\end{equation}
and $E_{k}\equiv\sqrt{k^2 + M^2}$, with $M$ the constituent quark mass.
(The equal-mass case is considered here; however, the extension to unequal
masses is straightforward.)  The quantum numbers $L,S$ range only over
the values allowed by $J^\pi$, and correspond to the usual
``nonrelativistic'' quantum numbers, while $\cal{L},\cal{S}$ can take
on all values allowed by the coupling to $J$, thus reflecting the role
of relativity in the calculation.  For $E > 0$, the amplitudes $b,d$
satisfy the RPA normalization condition \cite{ullah71}
\begin{equation}
  \sum_{L S} \int_{0}^{\infty} \frac{dk}{(2\pi)^3}
   \left [ {b}_{LSJ}^{2}(k) - {d}_{LSJ}^{2}(k) \right ] = 1 .
\label{rpanorm}
\end{equation}
Eqs.~(\ref{bklsj}) and~(\ref{dklsj}) are similar to the equations in
Ref.~\cite{piek92} for the two-fermion case.  This form is used for
convenience,
as the two can be related by charge conjugation.

\subsection{The RPA Equation}
Salpeter's equation can be cast in the following compact matrix form:
\begin{eqnarray}
  \left( \begin{array}{rr}
         H^{++}  &  H^{+-}  \\
         -H^{+-} &  -H^{++} \\
         \end{array}
  \right)
  \left( \begin{array}{c}
         {{B}} \\
         {{D}} \\
         \end{array}
  \right) =
E \left( \begin{array}{c}
         {{B}} \\
         {{D}} \\
         \end{array}
  \right) \;,
\label{RPA}
\end{eqnarray}
where the matrix elements of the ``Hamiltonian'' are given by
\begin{eqnarray}
  \langle k;LSJ | H^{++} |
    k^\prime ; L^\prime S^\prime J \rangle   &=&
  \langle k;LSJ | V^{++} |
  k^\prime ; L^\prime S^\prime J \rangle +
  2E_{k} (2\pi)^3 \delta(k- k^\prime)
  \delta_{L L^\prime}\delta_{S S^\prime} \;, \\
  \langle k;LSJ | H^{+-} |
   k^\prime ; L^\prime S^\prime J \rangle   &=&
  \langle k;LSJ | V^{+-} |
   k^\prime ; L^\prime S^\prime J \rangle    \;.
\end{eqnarray}
One recognizes that Salpeter's eigenvalue equation, as given by
Eq.~(\ref{RPA}), has the same algebraic structure as an RPA
equation~\cite{long84,resag93,parra94,piek92}.
Having identified the algebraic (RPA) structure of Salpeter's equation,
the same formalism developed by Thouless in his study of nuclear collective
excitations~\cite{thou60} will be employed.  Salpeter's, and in general any
RPA-like, equation can be rewritten as a Hermitian eigenvalue equation for
the square of the energy~\cite{ullah71,chi70}.  This implies that while the
square of the energy is guaranteed to be real, the energy itself might not.
The appearance of solutions having \mbox{$E^{2}$$<$0} signals, in the context
of nuclear collective excitations, an instability of the ground state against
the formation of particle-hole pairs --- a collective mode with imaginary
energy can build up indefinitely.  Thouless has shown that the stability
of the nuclear ground state depends on the Hermitian matrix
\begin{equation}
           \left( \begin{array}{cc}
                    H^{++} &  H^{+-}  \\
                    H^{+-} &  H^{++}  \\
                   \end{array}
            \right)
\end{equation}
being positive-definite---all its eigenvalues must be
greater than zero~\cite{thou60,ullah71,chi70}. This condition is
equivalent to requiring that both the sum and difference
matrices
\begin{eqnarray}
  H^+ &\equiv& \Big( H^{++} + H^{+-} \Big) \;,  \\
  H^- &\equiv& \Big( H^{++} - H^{+-} \Big) \;,
\end{eqnarray}
be positive-definite~\cite{chi70}. In this form, the stability
condition of Salpeter's equation is reduced to finding the
eigenvalues of the two Hermitian matrices $H^+$ and $H^-$. Thus,
the existence of a single negative eigenvalue, of either $H^{+}$ or
$H^{-}$, suffices to signal the instability.  It is this criterion that
was employed in Ref.~\cite{parra94} to examine the Lorentz structure of
the confining potential; also, it will be used in this work to set limits on
the strong coupling for the instantaneous OGE kernel in Salpeter's equation.
With these limits in hand, one is then able to carry out (see
Section~\ref{secspectra}) a study of the meson spectra in the framework of
Salpeter's equation.
\subsection{Numerical Solution of Salpeter's Equation}
Salpeter's equation is solved via expansion of the Salpeter amplitudes
in a suitable basis, thus enabling one to treat the instantaneous
confining and Coulomb kernels in configuration representation and the
relativistic kinetic energy operator in momentum representation, where
they are respectively local.  Here and in Ref.~\cite{parra94}, one uses
the radial eigenfunctions of the nonrelativistic harmonic oscillator,
$R_{nL}$, to expand the two amplitudes $B_{LSJ}$ and $D_{LSJ}$ in terms
of unknown coefficients
\begin{eqnarray}
  B_{LSJ}(k) &=& \sum_{n}^{n_{max}} B_{nLSJ} R_{nL}(k) \\
  D_{LSJ}(k) &=& \sum_{n}^{n_{max}} D_{nLSJ} R_{nL}(k) \;\; ,
\end{eqnarray}
up to \mbox{$n$=0,...,$n_{max}$} nodes in the basis, for a finite basis.
(Since the interaction is spherically symmetric, the magnetic quantum number
$M$ only denotes a $2J+1$-degeneracy and plays no dynamical role; hence
it can be dropped.)  This procedure results in a matrix equation for the
unknown coefficients $B_{nLSJ}$ and $D_{nLSJ}$ which can be diagonalized
using the method developed by Ullah and Rowe~\cite{ullah71}. Upon
diagonalization, one obtains $E^2$ and the (previously) unknown
coefficients from which one then can reconstruct the two amplitudes
$B_{LSJ}$ and $D_{LSJ}$, and, ultimately, the Salpeter wave function
$\chi^{E}$.

\section{Stability Analysis of Salpeter's Equation}
\label{secstability}
The stability analysis is performed by using potentials $V(r)$
having scalar, timelike-vector, and vector Lorentz structures
\begin{equation}
  V(r)\Gamma_1 \Gamma_2 = V(r)
    \cases{ {\bf 1}_{1}{\bf 1}_{2} \,,       & for scalar,   \cr
            \gamma^{0}_{1}\gamma^{0}_{2}\,,  & for timelike, \cr
            \gamma_{1}^{\mu}\gamma_{2\mu} \,, & for vector,  \cr}
\end{equation}
which are the relevant structures for the meson problem.
The analysis is concentrated on the pseudoscalar (\mbox{$J^{\pi}$=$0^{-}$})
channel; with \mbox{$L$=$S$=0}, this is the first
channel where the instability is likely to develop.  For this case, Salpeter's
equation, for the reduced amplitudes $b(k)\equiv kB(k)$ and
$d(k)\equiv kD(k)$, takes the following form
\begin{eqnarray}
  \left(+E - 2E_{k} \right) b(k)
   &=&
  \int_{0}^{\infty} \frac{d{k^\prime}}{(2\pi)^3}
    \biggl \{
       \langle k | {V}^{++} | k' \rangle b(k') +
       \langle k | {V}^{+-} | k' \rangle d(k')
     \biggl \} \;,
\label{bkpion} \\
  \left(-E - 2E_{k} \right) d(k)
   &=&
    \int_{0}^{\infty} \frac{d{k^\prime}}{(2\pi)^3}
     \biggl \{
       \langle k | {V}^{+-} | k' \rangle b(k') +
       \langle k | {V}^{++} | k' \rangle d(k')
     \biggl \} \;.
\label{dkpion}
\end{eqnarray}
In spite of the simplicity of the angular momentum content of this
channel, the matrix elements of the potential are complicated by
relativistic corrections. We define the angular-momentum components
of the potential by
\begin{equation}
  V_{\cal L}(k,k^\prime)
  = (4\pi)^{2}\int_{0}^{\infty} {dr} \;
    \hat{\jmath}_{\cal L}(kr) V(r) \hat{\jmath}_{\cal L}(k^\prime r) \;,
\label{vlkkp}
\end{equation}
with $\hat{\jmath}_{\cal L}(x) \equiv x j_{\cal L}(x)$ being the
Ricatti-Bessel function.  For scalar and timelike potentials, one has
\begin{eqnarray}
\lefteqn{\langle k | V^{++} | k' \rangle}\nonumber \\
 &=&
 \left( {E_{k}  + M \over 2 E_{k}  }  \right)
 \left( {E_{k'} + M \over 2 E_{k'} }  \right)
  \Bigg\{
   \left[ 1 + \zeta^{2}_{k}\zeta^{2}_{k'} \right] V_0(k,k') \mp
          2\zeta_{k}\zeta_{k'} V_1(k,k')
  \Bigg\}  \;,
  \label{vpionpp} \\
\lefteqn{\langle k | V^{+-} | k' \rangle }\nonumber \\
 &=&
 \left( {E_{k}  + M \over 2 E_{k}  }  \right)
 \left( {E_{k'} + M \over 2 E_{k'} }  \right)
  \Bigg\{
   \left[ \zeta^{2}_{k} + \zeta^{2}_{k'} \right] V_0(k,k') \pm
         2\zeta_{k}\zeta_{k'} V_1(k,k')
  \Bigg\}  \;,
  \label{vpionpm}
\end{eqnarray}
where the upper (lower) sign in the above expressions should be
used for scalar (timelike) potentials.  For vector potentials, one has
\begin{eqnarray}
\lefteqn{\langle k | V^{++} | k' \rangle }\nonumber \\
 &=&
 \left( {E_{k}  + M \over 2 E_{k}  }  \right)
 \left( {E_{k'} + M \over 2 E_{k'} }  \right)
  \Bigg\{
    \left[ 1 + \zeta^{2}_{k}\zeta^{2}_{k'} \right]
 + 3\left[ \zeta^{2}_{k} + \zeta^{2}_{k'}  \right]
  \Bigg\} V_0(k,k') \;,
  \label{vpionppvec} \\
\lefteqn{\langle k | V^{+-} | k' \rangle }\nonumber \\
 &=&
 \left( {E_{k}  + M \over 2 E_{k}  }  \right)
 \left( {E_{k'} + M \over 2 E_{k'} }  \right)
  \Bigg\{
    \left[ \zeta^{2}_{k} + \zeta^{2}_{k'}  \right]
  + 3\left[ 1 + \zeta^{2}_{k}\zeta^{2}_{k'} \right]
  \Bigg\} V_0(k,k') \;,
  \label{vpionpmvec}
\end{eqnarray}
where the kinematical variable
\begin{equation}
 \zeta_{k} \equiv {k \over E_{k} + M}
 \sim {\cal O}\left(\frac{k}{M}\right) \;\; ,
 \label{zeta}
\end{equation}
has been introduced to quantify the importance of relativity.  In particular,
for scalar and timelike potentials, relativistic corrections arising from the
mixing of positive and negative energy states (as characterized by
$\langle k | {V}^{+-} | k' \rangle$) appear as
${\cal O}\left(\frac{k^2}{M^2}\right)$ relative to the unmixed (Breit) case.
This contrasts with the behavior for vector potentials, where both matrix
elements contain ${\cal O}\left(1\right)$ terms, implying that the contribution
of negative-energy states for vector potentials will impact results more than
the scalar and timelike cases.  This can be seen by taking
the nonrelativistic limit
($ \zeta_{k},\zeta_{k'} \rightarrow 0$):
\begin{eqnarray}
 \langle k | V^{++} | k' \rangle & \rightarrow &
   \cases { V_0(k,k')~,     & for scalar and timelike, \cr
            V_0(k,k')~,     & for vector,
          } \\
 \langle k | V^{+-} | k' \rangle & \rightarrow &
   \cases { 0~,             & for scalar and timelike, \cr
            3 V_0(k,k')~,   & for vector.
          }
\end{eqnarray}
The ${\cal O}\left(1\right)$ term in $V^{+-}$ in the vector case stems
from the additional spacelike (${\bf \gamma}_{1}\cdot{\bf \gamma}_{2}$)
contribution relative to the scalar and timelike cases, contrary to the usual
assumption that ${\bf \gamma}_{1}\cdot{\bf \gamma}_{2}$ induces
${\cal O}\left(\frac{k^2}{M^2}\right)$ corrections to the nonrelativistic
potential.  One should also note that the
${\bf \gamma}_{1}\cdot{\bf \gamma}_{2}$ contribution mixes upper and lower
components via a spin-spin term, thus giving rise to much stronger splittings
than either the scalar or timelike cases.

\subsection{Stability Analysis for the Confining Kernel}
In this subsection, the stability analysis carried out in
Ref.~\cite{parra94} is reviewed for completeness.  For confining
potentials, the Fourier transform [Eq.~(\ref{vlkkp})] is
ill-defined.  Hence, in examining confinement in momentum space in
Ref.~\cite{parra94}, the following regularization for the spatial part
of the potential was employed~\cite{spence86,maung93}:
\begin{equation}
  V(r) = \sigma r e^{-\eta r} \equiv
         \sigma \frac{\partial^{2}}{\partial\eta^{2}}
         \frac{e^{-\eta r}}{r}  \;.
\end{equation}
The Fourier transform of the potential is now well behaved and is given by
\begin{equation}
    V({\bf k}-{\bf k}') =
    \frac{\partial^{2}}{\partial\eta^{2}}
    \left[
      \frac{4\pi\sigma}{({\bf k}-{\bf k}')^2 + \eta^2}
    \right] \;.
\label{vkansatz}
\end{equation}
Evidently, one is interested in studying the stability of Salpeter's
equation in the limit of $\eta \rightarrow 0$.  The stability analysis
required the explicit evaluation of $V^{+}$ and $V^{-}$. These were
computed with the help of Eqs.~(\ref{vpionpp}) and~(\ref{vpionpm})
\begin{eqnarray}
  V^{+}(k,k') &\equiv&
  \langle k | V^{++} + V^{+-} | k' \rangle = V_0(k,k') \;,
 \label{vplus} \\
  V^{-}(k,k') &\equiv&
  \langle k | V^{++} - V^{+-} | k' \rangle = V_0(k,k')\xi(k,k') \;,
 \label{vminus}
\end{eqnarray}
where one introduced relativistic ``correction'' factors, separately,
for scalar and timelike confinement
\begin{eqnarray}
 \xi_{\rm s}(k,k') &\equiv&
  \left[
   {M^{2} \over E_{k}E_{k'}} -
   {kk'   \over E_{k}E_{k'}}
   {V_1(k,k') \over V_0(k,k')}
  \right] \;,
 \label{xis} \\
 \xi_{\rm v}(k,k') &\equiv&
  \left[
   {M^{2} \over E_{k}E_{k'}} +
   {kk'   \over E_{k}E_{k'}}
   {V_1(k,k') \over V_0(k,k')}
  \right] \;.
 \label{xiv}
\end{eqnarray}
For both scalar and timelike confining kernels, $H^{+}$ remained
positive-definite, while $H^{-}$ was positive-definite only for
timelike confinement, and not for scalar confinement; a scalar
confining kernel in Salpeter's equation leads to imaginary-energy
solutions---irrespective of the constituent quark mass~\cite{parra94}.
Also, if a mixture of scalar and timelike structure for the Lorentz
structure of the potential is employed
\begin{equation}
  \Gamma \equiv x\gamma^{0}_{1}\gamma^{0}_{2} +
                (1-x) {\bf 1}_{1}{\bf 1}_{2}  \;,
\end{equation}
where $x$ denotes the fraction of timelike structure, then
\begin{eqnarray}
  V^{+}(k,k') &=& V_0(k,k') \;,  \\
  V^{-}(k,k') &=& V_0(k,k')
  \left[
         {M^{2} \over E_{k}E_{k'}} +
   (2x-1){kk'   \over E_{k}E_{k'}}
   {V_1(k,k') \over V_0(k,k')}
  \right] \;.
\end{eqnarray}
Again, $H^{+}$ remains positive definite, while in contrast,
$H^{-}$ is positive definite only for $x\ge 1/2$. Hence, any mix of
scalar and timelike Lorentz structures has stable solutions only for
$x$ in the interval $0.5\le x\le 1$.  This fact will become important
in the study of the meson spectra.

\subsection{Stability Analysis for the Instantaneous OGE Kernel}
\label{seccoulomb}
One now considers the (short-range) OGE part of the kernel.  The
stability analysis is performed for a pure Coulomb potential, for
both timelike and vector Lorentz structures.  (Although an OGE
kernel of vector Lorentz character will be employed for the spectra
analysis, the timelike results are presented here for comparison.)
The spatial part of the (instantaneous) Salpeter kernel is
\begin{equation}
  V_{OGE}(r) = \frac{-\alpha_{s}}{r}
\label{vcoul}
\end{equation}
with the Fourier transform of $V_{OGE}(r)$ given by
\begin{equation}
    V_{OGE}(|{\bf k}-{\bf k}'|)
  = \lim_{\eta\rightarrow 0}
    \left[
      \frac{-4\pi\alpha_{s}}{({\bf k}-{\bf k}')^2 + \eta^2}
    \right] \;.
\label{vkkpcoulomb}
\end{equation}
where $\alpha_{s}$ is the strong coupling (taken here to be independent
of the quark-antiquark separation $r$, or equivalently, of the momentum
transfer $Q$).

For the timelike case, one has essentially carried out all necessary
manipulations in Ref.~\cite{parra94}.  Eqs.~(\ref{vplus})
and~(\ref{vminus}) are completely general for any potential with timelike
structure, and one only has to calculate the necessary multipoles of the
potential in momentum space to complete the analysis.  One looks to the
$k=k^\prime$ limit for which the Coulomb singularity structure is
manifest. For $V_0$ and $V_1$, one finds (in the $\eta \rightarrow 0$
limit):\footnote{The OGE subscript will be discarded, and future references
to $V$ will be taken as indicating the Coulomb potential, unless otherwise
noted.}
\begin{eqnarray}
  V_{0}(k,k'=k) &=& -8\pi^{2}\alpha_s \frac{1}{2}
                    \ln\left(\frac{4 k^2}{\eta^2}\right) +
                   {\cal O}(\frac{\eta^2}{k^2}) \;, \\
  \label{vtcoulzero}
  V_{1}(k,k'=k) &=& -8\pi^{2}\alpha_s
                    \frac{1}{2}\left[ \ln\left(\frac{4 k^2}{\eta^2}\right) - 1
                               + {\cal O}(\frac{\eta^2}{k^2})
                               \right]  \;.
  \label{vtcoulone}
\end{eqnarray}
with the leading singularities cancelling in forming the ratio
\begin{equation}
  \lim_{\eta\rightarrow 0}
  {V_{1}(k,k'=k) \over V_{0}(k,k'=k)} =
  \lim_{\eta\rightarrow 0}
  \left[1 - \frac{1}{\ln\left({\eta^{2} \over 4k^{2}}\right)} +
        {\cal O}(\eta^2/k^2)\right] = 1 \;.
 \label{coulimit}
\end{equation}
and one sees that $\xi_{\rm v}(k,k)\rightarrow 1$ again.  However, in
contrast to the confining potential, the stability of the RPA matrix
in present case is not assured, as the ``nonrelativistic''
equation for $H^{+}$ can become unbounded from below
for sufficiently large $\alpha_s$. This implies the existence of an
upper limit on the effective strong coupling that one can use for a
timelike Coulomb potential in Salpeter's equation, which affects the
determination of wave functions and spectra within the model.  Thus,
it becomes necessary to determine an upper limit on $\alpha_s$
in order to avoid the instability.  Before doing this, the vector case will be
considered to see if the same problem exists.  The matrix elements of the
potential are somewhat complex; however, $V^{-}$ and $V^{+}$ are quite simple:
\begin{eqnarray}
  V^{+} &=& 4 V_{0}(k,k^\prime) ,\\
 \label{vminuscouv}
  V^{-} &=& -2 \left[\frac{M^2}{E_{k} E_{k^\prime}}\right] V_{0}(k,k^\prime) .
 \label{vpluscouv}
\end{eqnarray}
Since the minus sign in $V_{0}$ leaves $V^{-}$ positive, one sees that
$H^{-}$ is actually positive-definite, so is bounded from below, and one can
concentrate on $H^{+}$.\footnote{In fact, since $V^{-}$ is repulsive, $H^{-}$
does not even support bound states.}  The eigenvalue equation for $H^{+}$ is
again the ``nonrelativistic'' Schr\"odinger problem, but with a
strong coupling four times the original $\alpha_s$:
$$
  \alpha \equiv 4\alpha_s  .
$$
This is a reflection of the strong upper-to-lower coupling induced by the
spacelike component of the vector Lorentz structure.
Hence, for both timelike and vector cases, one has to determine an upper limit
for $\alpha_s$ at which Salpeter's equation becomes unbounded, noting that the
upper limit for the vector case will be one-fourth that in the timelike case.

\subsection{Determination of the Ground State Energy at the
            ``Critical Coupling''}
Evidently, one is interested in the spectrum of the
``semirelativistic'' Coulomb Hamiltonian
\begin{equation}
  H = 2\sqrt{k^2 + M^2}  - {\alpha \over r}~;~~ \alpha > 0 \;.
  \label{eq.semirel_hamil}
\end{equation}
As we mentioned in Sec.~\ref{seccoulomb}, the existence of a single
negative eigenvalue,
of either $H^{+}$ or $H^{-}$, suffices to signal the instability.
If the Hamiltonian (\ref{eq.semirel_hamil}) is unbounded from below, or is
bounded but has at least one negative eigenvalue, then the RPA instability
develops. In 1977, Herbst\cite{herbst77} was able to show that: (a) If
$\alpha > \alpha_c \equiv 4/\pi$, then the semirelativistic Coulomb
Hamiltonian is unbounded from below; (b) If $\alpha \le \alpha_c$, then
all eigenvalues are greater than or equal to $0$. Specifically, Herbst
showed that
\begin{equation}
  E \ge 2M \sqrt{ 1 - \left({\alpha\over\alpha_c}\right)^2 }~\;;
  \quad {\rm for \;} \alpha \le \alpha_c \;.
  \label{herbst}
\end{equation}
However, for the RPA equations to be stable one must show that
all eigenvalues must be positive---a result that does not follow
from Eq.~(\ref{herbst}) at the critical coupling. Although our primary
interest in the spectrum of the semirelativistic
Coulomb Hamiltonian stems from the stability analysis of Salpeter's
equation, the semirelativistic Coulomb problem is still of considerable
theoretical interest\cite{raynal94,lucha94}. Thus, in this subsection we
present a variational analysis that, to our knowledge, represents the
best estimate of the ground-state energy available in the literature.

To date, no analytic solution for the ground-state energy of the
semirelativistic Coulomb problem [Eq.~(\ref{eq.semirel_hamil})]
exists. Thus, we determine an upper bound for the ground-state
energy ($E_{0}$) by using a Rayleigh-Ritz variational method.
That is, given a trial wave function $|\Psi\rangle$, an upper bound
to the ground-state energy is given by the expectation value of $H$:
\begin{equation}
  E_{0} \le
  \langle H \rangle \equiv \langle \Psi | H | \Psi \rangle \;;
  \quad \langle \Psi | \Psi \rangle \equiv 1 \;.
    \label{eq.expe_value}
\end{equation}
In practice, one attempts to include the relevant physics by including
a set of variational parameters into $|\Psi\rangle$. Then, one
minimizes Eq.~(\ref{eq.expe_value}) with respect to those parameters
to determine the variational bound and the ``optimal'' $|\Psi\rangle$.
In configuration space, the variational wave function is taken to be
\begin{eqnarray}
  \Psi({\bf r}) & = & \sqrt{1 \over 4\pi} R(r)  \;, \\
   R(r)         & = & N_R {r^\epsilon \over r} e^{-\gamma r}
                      \Big[1 + c_1 (\gamma r)
                      + c_2 (\gamma r)^2 + \cdots\Big]~,
                      \label{eq.expansion}
\end{eqnarray}
where the normalization constant is given by
\begin{equation}
  N_R = \left[ (2\gamma)^{1+2\epsilon} \over
         {\displaystyle\sum_{n,m=0}^{\infty}
         { c_{n-m} c_m \over 2^n} \Gamma(1+n+2\epsilon) } \right]^{1/2}~.
\end{equation}
Note that this form of the wave function is appropriate for the
$L=0$ channel---where the instability should first develop. The
expansion for the variational wave function is complete and its
form is motivated by the analytic solution of the Dirac-Coulomb
problem\cite{sakurai73}. In particular, in the weak-coupling limit
($\alpha\ll 1$) one recovers the nonrelativistic result by
choosing $\epsilon=1$, $\gamma=M\alpha/2$, and $c_1=c_2=\cdots=0$,
i.e.,
\begin{eqnarray}
  && E = 2M ( 1 - {\alpha^2 \over 8} ) \\
  && R(r) = \sqrt{M^3 \alpha^3 \over 2} e^{-{M\alpha r/2}}~.
\end{eqnarray}
In contrast, in the strong-coupling limit ($\alpha \sim \alpha_{c}$)
one expects that the wave function will become localized near the
origin, as the energy can benefit from the strong Coulomb attraction.
Indeed, the asymptotic behavior of the wave function near the origin is
known analytically\cite{durand83,friar84}
\begin{equation}
   {\tan\left({\pi\over2}\epsilon\right) \over
    \left({\pi\over2}\epsilon\right)} =
   {\alpha_{c} \over \alpha} \;.
\end{equation}
Note that $\epsilon \rightarrow 0$ as $\alpha$ approaches the
critical coupling.
The variational wave function has also an analytic representation
in momentum space. That is,
\begin{eqnarray}
  \Psi({\bf k}) & = & \sqrt{1 \over 4\pi} R(k) \;,  \\
   R(k)         & = & {1\over k} \sqrt{2\over\pi} N_R \sum_{n=0}^{\infty}
                      c_n \gamma^{n}
                      {\Gamma(1+n+\epsilon) \over
                      (k^2+\gamma^2)^{(1+n+\epsilon)/2} }
                      \sin\left[ (1+n+\epsilon) \tan^{-1}\left({k\over\gamma}
                      \right) \right] \;.
  \label{eq.pexpansion}
\end{eqnarray}
One should note that since the constituent quark mass is the
only dimensionful parameter in the problem, the dimensionless
ratio ${\langle H \rangle / M}$ is a function of only the
coupling constant $\alpha$. In what follows, all expectation
values will be written in units of the constituent mass $M$ and
expressed in terms of the dimensionless parameter $a\equiv\gamma/M$.

For values of $\alpha$ not too close to the critical coupling ($\alpha
{\hbox{${\lower.40ex\hbox{$<$}\atop\raise.20ex\hbox{$\sim$}}$}}1.25
< \alpha_{c} \approx 1.273$) the minimization procedure for
$\langle H \rangle / M$ is straightforward and yields the variational
energy and optimal parameters ($a$ and $\epsilon$) that are displayed in
Table~\ref{table4chap3}. Note, the table reflects the appropriate
parameters for the case $c_{1}=c_{2}=\cdots=0$. In contrast, the
minimization procedure is highly nontrivial near the critical coupling
(and, thus, for $\epsilon \rightarrow 0$) as, both, the kinetic and
potential energy diverge as $1/\epsilon$ (see Appendix). Hence, in
order to compute the variational energy for small values of $\epsilon$
we perform a Laurent expansion of the expectation value of $H$,
i.e.,
\begin{equation}
  {\langle H \rangle \over M} = {h_{-1} \over \epsilon} + h_{0}
                                + h_{1} \epsilon + \cdots~~.
\end{equation}
The first term, $h_{-1}$, must vanish faster than
$\epsilon \sim (\alpha_c-\alpha)^{1/2}$ as $\alpha$
approaches the critical coupling; otherwise one would contradict
Herbst findings. Indeed, we can extract the critical coupling
($\alpha_c=4/\pi$) by demanding that the coefficient of the singular
term in the series vanishes. Note, we have shown in the appendix that
$h_{-1}$ vanishes as $(\alpha_c-\alpha)$ in the $c_{1}=c_{2}=\cdots=0$
limit (see Eq.~\ref{eq.h_expt}).

In principle, we could find the variational energy in the
$\alpha=\alpha_c$ limit by minimizing $h_0$ using the trial
wave function (\ref{eq.expansion}). In practice, however,
we can manage only a small number of variational parameters.
Thus, we proceed by, first, minimizing the expectation value
of $H$ using only one term in the polynomial expansion in
Eq.~(\ref{eq.expansion}), i.e., we set $c_1=c_2=\cdots=0$.
Note, most of the details related to this minimization procedure
are presented in the appendix. Next, we compute the variational
energy by using a two-parameter ($a$ and $c_1$) wave function.
This procedure is straightforward but tedious. However, it
enables us to determine the importance of higher-order
``corrections'' in the polynomial expansion as well as the
rate of convergence to the ground-state energy. Our results
are summarized below:
\begin{equation}
  \displaystyle{\langle H \rangle \over M} =
  \cases{0.968583, & \ for $a=0.7926$ and $c_{1}=0.0000$; \cr
         0.968514, & \ for $a=0.9359$ and $c_{1}=0.1779$.}
\end{equation}
Moreover, an initial study with three variational parameters
($a$, $c_1$, and $c_2$) suggests that the two-parameter energy is
accurate to, at least, one part per million. To our knowledge this
represents the most accurate value for the ground-state energy of
the semirelativistic Coulomb problem presented to date. In this way,
our small contribution to Herbst work reads: If $\alpha \le \alpha_c=
4/\pi$, then all eigenvalues (in units of $M$) are greater than or
equal to 0.968514. Note, our results are consistent with those
presented by Raynal and collaborators in a comprehensive study of the
semirelativistic Coulomb problem\cite{raynal94}. Their analysis sets
lower and upper bounds---differing by less than 1\%---for the
ground-state energy ($0.9650 \le E_{0} \le 0.9686$) at $\alpha=\alpha_c$.

The remaining question to be answered is how, if at all, does
the presence of the linear confining potential alter this
stability analysis?
The answer is that only the value of the finite piece is changed,
at $\alpha=\alpha_{c}$. An explicit calculation gives for the
expectation value of the confining potential
(with $c_1=c_2=\cdots=0$)
\begin{equation}
  {\langle V \rangle \over M}  =
  {\sigma \over 2 a M^2} + {\sigma \over a M^2} \epsilon \;.
\end{equation}
Thus, there is no ${\cal O}(\epsilon^{-1})$ contribution to
the expectation value of $\langle H \rangle$ coming from the
confining potential; only a positive contribution,
${\sigma / 2 a M^2}$, remains at $\epsilon\rightarrow 0$.
Note that, because of the additional dimensionful parameter
$\sigma$, the contribution from the confining potential,
unlike the Coulomb contribution, depends on the value of
the constituent mass. Hence, for a Coulomb potential of the
form of Eq.~(\ref{vcoul}), stability of Salpeter's equation is achieved
by demanding that
\begin{equation}
  \alpha_{s} \le \cases{
    \displaystyle{4 / \pi} \approx 1.273~, & for timelike; \cr
    \displaystyle{1 / \pi} \approx 0.318~, & for vector,   \cr}
\end{equation}
independent of the constituent quark mass.

\section{Meson Spectra}
\label{secspectra}
The heavy quarkonia ($c\bar{c}$ and $b\bar{b}$) and the light quarkonia
($u\bar{u}$ and $s\bar{s}$) are investigated in the instantaneous
Bethe-Salpeter framework.  The analysis is intended to be qualitative
in nature, with regard to various effects stemming from the instantaneous
approximation; hence a simple generalization of the Cornell
potential~\cite{eichten7880} is employed.  The virtue of this approach
is its simplicity.  The few parameters of the model, constrained by the
stability analysis, will emerge from the ``best'' fit to the meson spectra.
One wishes to reproduce the interesting physics (e.g.~hyperfine structure)
through relativistic effects, rather than by a fine tuning of many
parameters.

For the heavy quarkonia, the mass spectra are reasonably described, but
details such as the fine and hyperfine structure are less so.  For the
light quarkonia, the mass spectra are less reasonably described: the
pion cannot be accurately modelled within this framework without losing
the remaining spectra. At the very least, a more sophisticated
phenomenology will be required to accurately describe static properties
of the heavy and light quarkonia.

For the remainder of the section, the following program is carried out: the
various models to be considered within the Salpeter framework are defined.
Then, phenomenological fits to the experimental spectra for heavy mesons are
carried out, and a quantitative analysis of the various approximations in the
Salpeter framework and their effect on the spectra and splittings is
performed. The latter half of the section is concerned with the light mesons.
\subsection{Fits to Heavy Mesons}
\subsubsection{Form of the Interaction}
The spatial part of the potential is based on the Cornell potential;
that is, confinement is parameterized by a linear potential plus a constant,
and the asymptotically-free regime is parameterized by an instantaneous
OGE potential.  Since a good description of the mass spectra of the heavy
quarkonia can be obtained with a linear confining potential plus the
nonrelativistic reduction of the OGE piece, this is a natural first choice
for the Salpeter equation.  (One can, of course, generalize the OGE piece
to include running coupling effects motivated from perturbative QCD.)

A mixture of scalar and timelike Lorentz structures for the confining
kernel is considered, as a number of authors~\cite{gara89,lagae92} have
suggested that an admixture of scalar and timelike confinement is
necessary to reproduce the experimental spectra and splittings.
The full vector structure is not incorporated in the confining kernel,
as this leads to an instability similar to the scalar case.  The full
vector structure in the Feynman gauge is used for the OGE kernel.
Thus, the instantaneous Bethe-Salpeter kernel is parameterized as
\begin{equation}
  V(r)\Gamma_{1,2} =
  \left[\sigma r + c_{\sigma}\right]
  \left[ x_{\sigma}\gamma^{0}_{1}\gamma^{0}_{2} +
  (1 - x_{\sigma}){\bf 1}_{1}{\bf 1}_{2}\right]
 -\left[\frac{\alpha_{s}}{r}\right]\gamma^{\mu}_{1}\gamma_{2\mu}
\end{equation}
where $\sigma$ is the confining string tension, $x_{\sigma}$ controls
the mixing between scalar and
timelike confinement, and $\alpha_s$ is the (scale-independent) strong
coupling constant.  One should note that $\sigma$, $c_{\sigma}$, $x_{\sigma}$,
$\alpha_{s}$, and the quark mass are the only ``free'' parameters in the
approach.  From the results of the stability analysis, $x_{\sigma}$ and
$\alpha_s$ are restricted to the following values in the Salpeter model
in order to have real eigenvalues:
$$
  0.5 \leq x_{\sigma} \leq 1.0 \;\;\;\; ; \;\;\;\;
  \alpha_s \le \frac{1}{\pi} \approx 0.318 \;\; .
$$
Recall that a typical value for $\alpha_s$ is 0.24\cite{spence93}.
One should note that the Breit ($V^{+-}\equiv 0$) model has no such
restrictions in principle, saving the possibility of the spectrum becoming
unbound from below if the strong coupling becomes too large, similar to the
problem encountered for the relativistic Coulomb problem of
Section~\ref{secstability}.  However, for comparative purposes,
$x_{\sigma}$ is restricted to the same range in both models.

\subsubsection{Data and Procedures}
One solves for eigenenergies of the Breit (no coupling between positive
and negative energy states) and full Salpeter equations, expanding the
Salpeter partial-wave amplitudes in an oscillator basis of 20 states
per partial wave (\mbox{$n_{max}$=20}, or up to 19 nodes in the amplitudes)
to insure adequate convergence of the solutions, and with an oscillator
parameter value \mbox{$\beta$=0.6 GeV} suitable for the heavy mesons.
Initially, there are six free parameters in the model: $\sigma$,
$\alpha_{s}$, $M_{c}$, $M_{b}$, $x_{\sigma}$, and $c_{\sigma}$.
The same $\sigma$, $x_{\sigma}$, and $\alpha_{s}$ are used to
fit both charm and beauty.  $c_{\sigma}$ is set to zero, assuming
the long-range part of the kernel to be less important for the
heavy mesons.  The scalar-timelike mixing parameter $x_{\sigma}$ will be
allowed to take only the values 0.5 and 1.0, corresponding to
equally-mixed scalar and timelike structure and pure timelike structure,
respectively.  Hence, the model for the heavy mesons will only have
four free parameters, which are determined through minimization of the
$\chi^2$ function
\begin{equation}
  \chi^{2}(\sigma,\alpha_{s},M_{c},M_{b})
  = \sum_{i=1}^{N} \frac{(E_{ith}-E_{iexp})^2}{\sigma_{ith}^2 +
    \sigma_{iexp}^2}
\end{equation}
using a nonlinear optimization routine~\cite{minuit92}.  The $E_{iexp}$ and
$\sigma_{iexp}$ are the experimental masses and associated errors chosen
for the fit, here the first two observed $1^{-}$ states of b-quarkonium,
the first two observed $0^{-}$ states and the first five observed
$1^{-}$ states of charmonium.  These states are used to determine a
set of parameters for each choice of Lorentz structure of the confinement.
Then, one observes how each of the corresponding spectra agrees with the
experimental data overall.  The $E_{ith}$ and $\sigma_{ith}$ are the
calculated masses and their errors within the model (which were taken
to be 5-10 MeV).  The incorporation of a theoretical error allows one,
in principle, to ``force'' a better fit to some states, at the risk of
possibly degrading the fit with respect to the rest of the spectra;
however, all states were weighted equally in this regard.

\begin{figure}[tb]
\centering
\epsfig{file=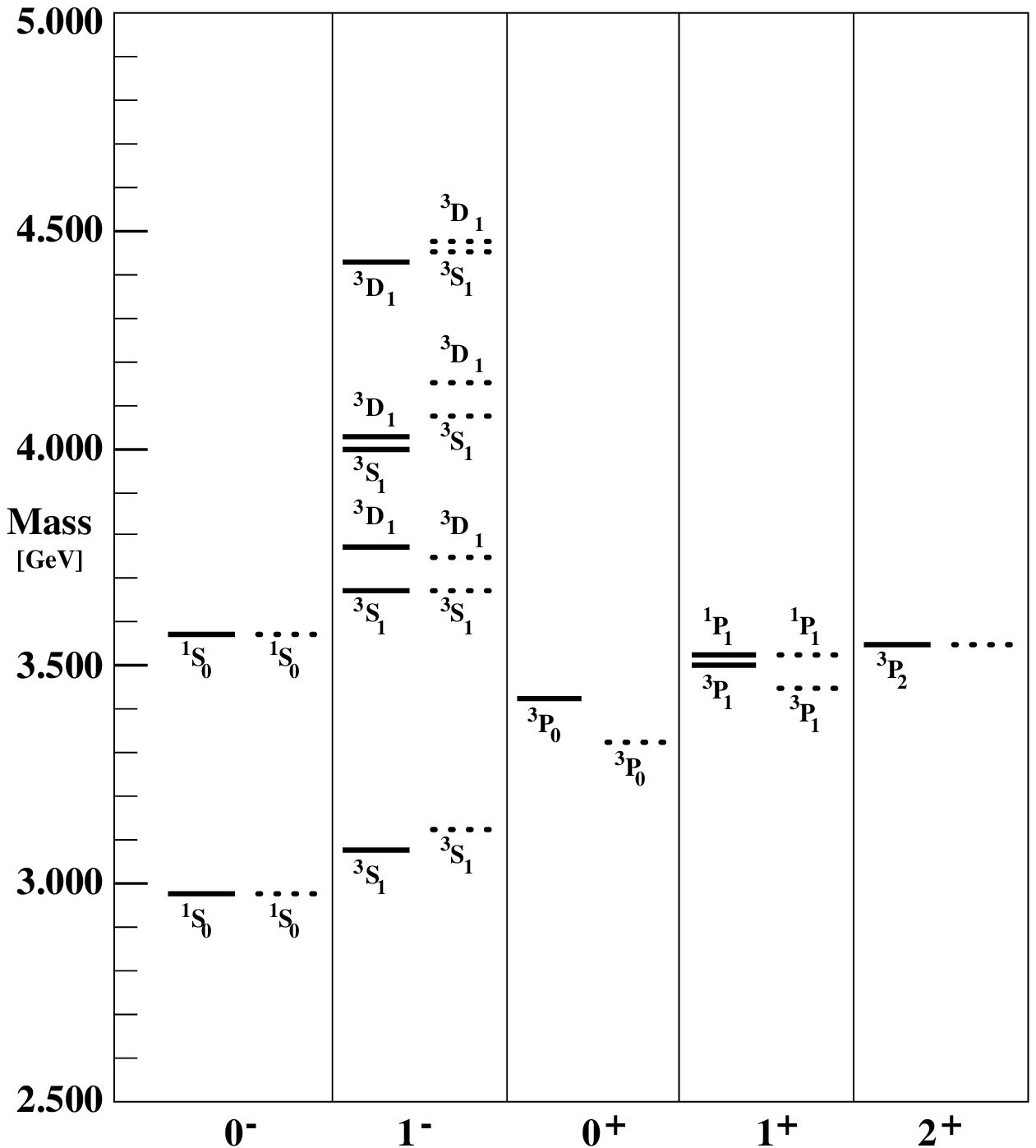}
\caption{Charmonium mass spectrum for Breit, with timelike confinement
              and a vector Coulomb contribution. The experimental
              numbers are in the left-hand column for each spin-parity.
              The spectroscopic notation for coupled states is that
              of the leading component in the calculation.}
\label{fig2chap4}
\end{figure}

\begin{figure}[tb]
\centering
\epsfig{file=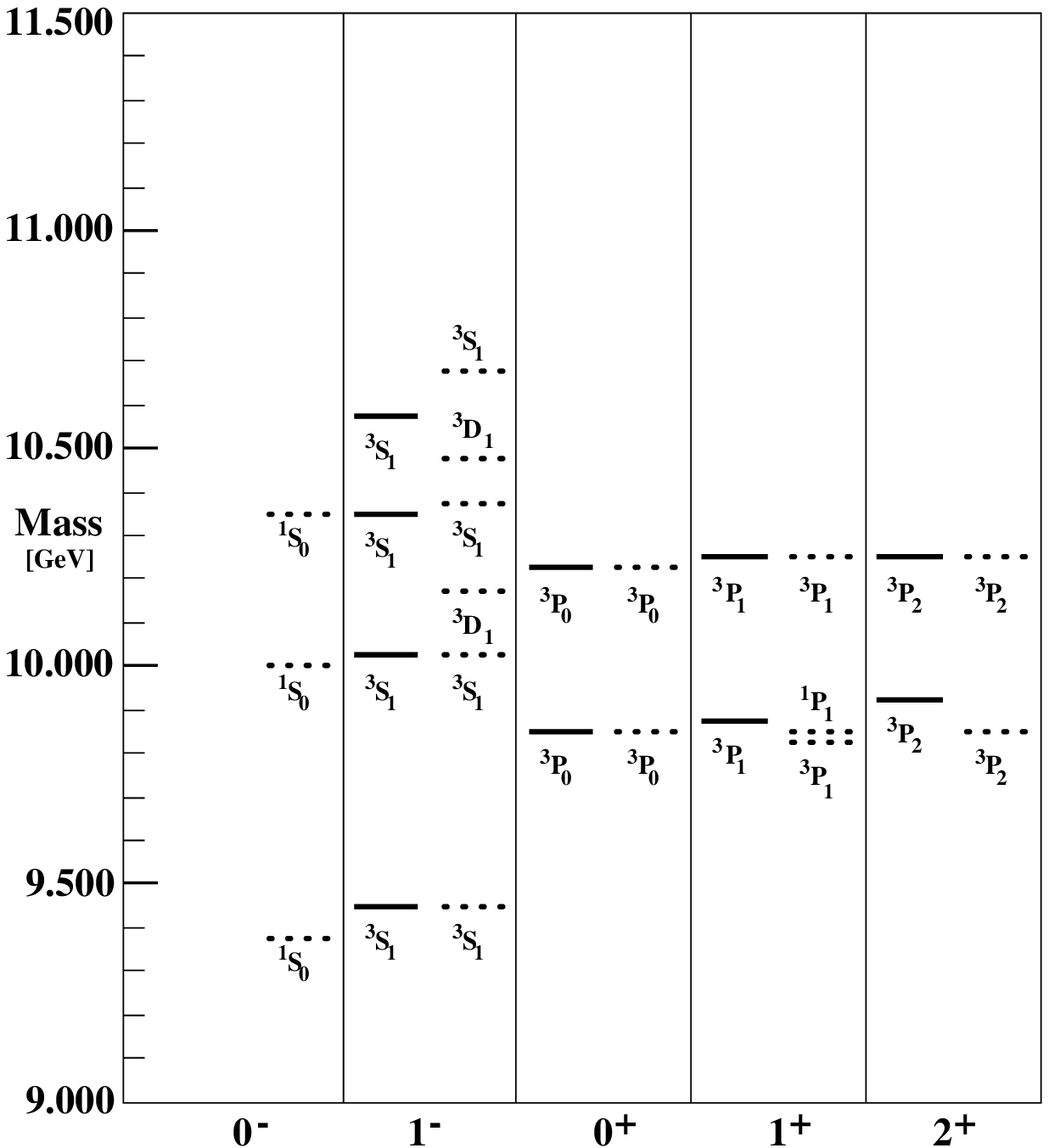}
\caption{Beauty quarkonium mass spectrum for Breit, with timelike
         confinement and a vector Coulomb contribution.  The experimental
         numbers are in the left-hand column for each spin-parity.
         The spectroscopic notation for coupled states is that of the
         leading component in the calculation.}
\label{fig6chap4}
\end{figure}

The minimization results for the four parameters for the four models
(see Table~\ref{table1achap4} for the appropriate definitions) are
summarized in Table~\ref{table1chap4}, while the charm and beauty
quarkonia spectra calculated for each model are summarized in
Table~\ref{table2chap4} and Table~\ref{table3chap4}. The experimental
masses were taken from the Review of Particle Properties~\cite{hikasa92}
(with the exception of the $h_{c}$, which was taken from
reference~\cite{armstrong92}).  Graphical depictions of the spectra for
each $J^{\pi}$ channel are shown for the Breit-Timelike model in
Figures~\ref{fig2chap4}~and~\ref{fig6chap4}.  $J^{\pi}$ states are
named by their main $^{2S+1}L_{J}$ components.  One notes that for
all cases, the parameters fall within commonly accepted ranges for
quark potential models.

\begin{figure}[tb]
\centering
\epsfig{file=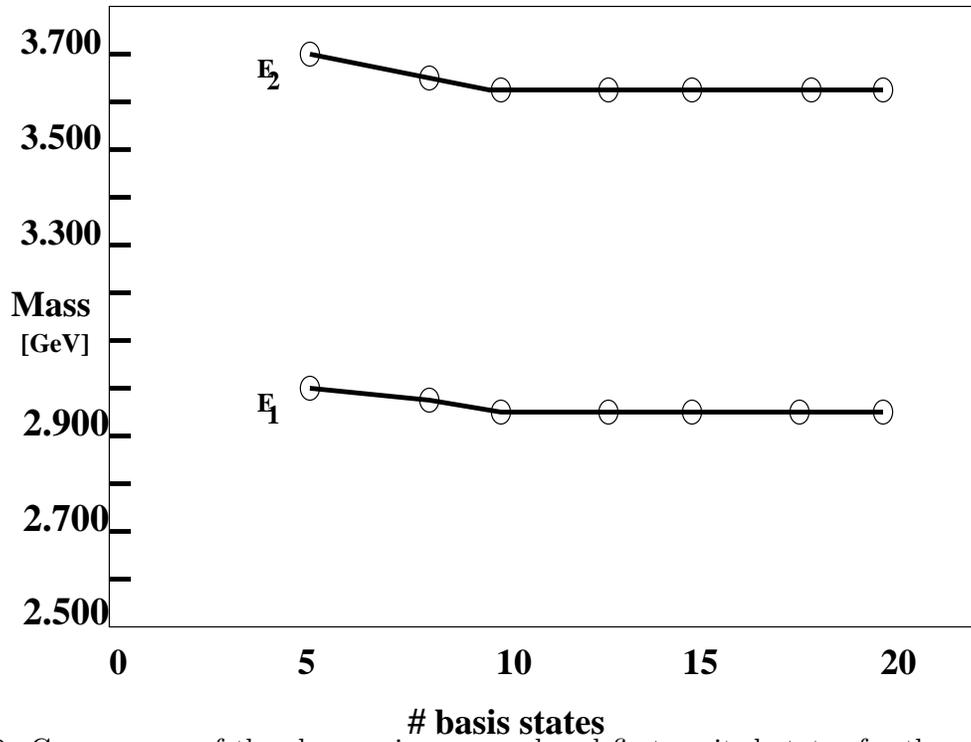}
\caption{Convergence of the charmonium ground and first excited states
         for the pseudoscalar channel as the number of states in the
         oscillator basis is increased, for Salpeter with mixed
         confinement as in Table~\protect\ref{table1chap4}. The basis
         parameter is \mbox{$\beta$=0.6} GeV for all cases.}
\label{fig9chap4}
\end{figure}

Figure~\ref{fig9chap4} shows the convergence of the ground and
first-excited state energies as the number of states in the basis
is increased; adequate convergence is achieved with 20 states in
the basis.  The numbers quoted for the higher-lying states in
Table~\ref{table2chap4} and Table~\ref{table3chap4} should be noted
with caution, as typically a 10-20 MeV shift in the energies for
the fourth-to-fifth eigenstates in going from 18 to 20 basis states
is encountered.  In particular, the larger differences in the table
with respect to the experimental data for the higher excited states
are indicative of an insufficient number of states in the basis.
Also, all the models appear to be somewhat deficient in comparison
to results given by Long~\cite{long84} and Spence and
Vary~\cite{spence93}, which examine both Breit and Salpeter
equations for a scalar confining kernel and a vector Coulomb kernel.
However, those studies allowed more freedom in determining the Salpeter
solutions.  Long utilizes an oscillator basis but minimizes each
eigenstate separately; each state has a different value of $\beta$
characterizing it, rather than one value for all states.  While this
procedure does minimize the eigenenergies with fewer states in the
basis, the disadvantage is that the eigenstates are not orthogonal,
which would be a problem in the calculation of matrix elements, and
in ensuring the proper normalization of bound states.  Spence and
Vary use a spline basis~\cite{deboor78} as well as an additional
interaction (a so-called ``Breit'' interaction) which makes comparison
more difficult.  However, in their work, solutions with imaginary
roots (for the light mesons in particular) were discarded on the
claim that the imaginary roots appear far from the real roots of
interest in the complex plane when the spline basis is chosen as in the
study~\cite{spence93}---a procedure that yields (apparently) stable
solutions of the Salpeter equation for scalar confinement, in contrast
to the results presented here.  This latter approach differs
drastically from the viewpoint adopted in this work, which is that the
onset of imaginary solutions should indicate that a particular interaction
is physically inappropriate within the model.  (It is rather amusing to note
that, when examining a kernel that can lead to instabilities, one can
``tune'' the oscillator basis to get apparently stable solutions, but
that either shifting the value of $\beta$ or increasing the number of
states in the basis (or both) reveals the instability.)

\subsubsection{Comparison of Approximations to Salpeter's Equation}
\begin{figure}[tb]
\centering
\epsfig{file=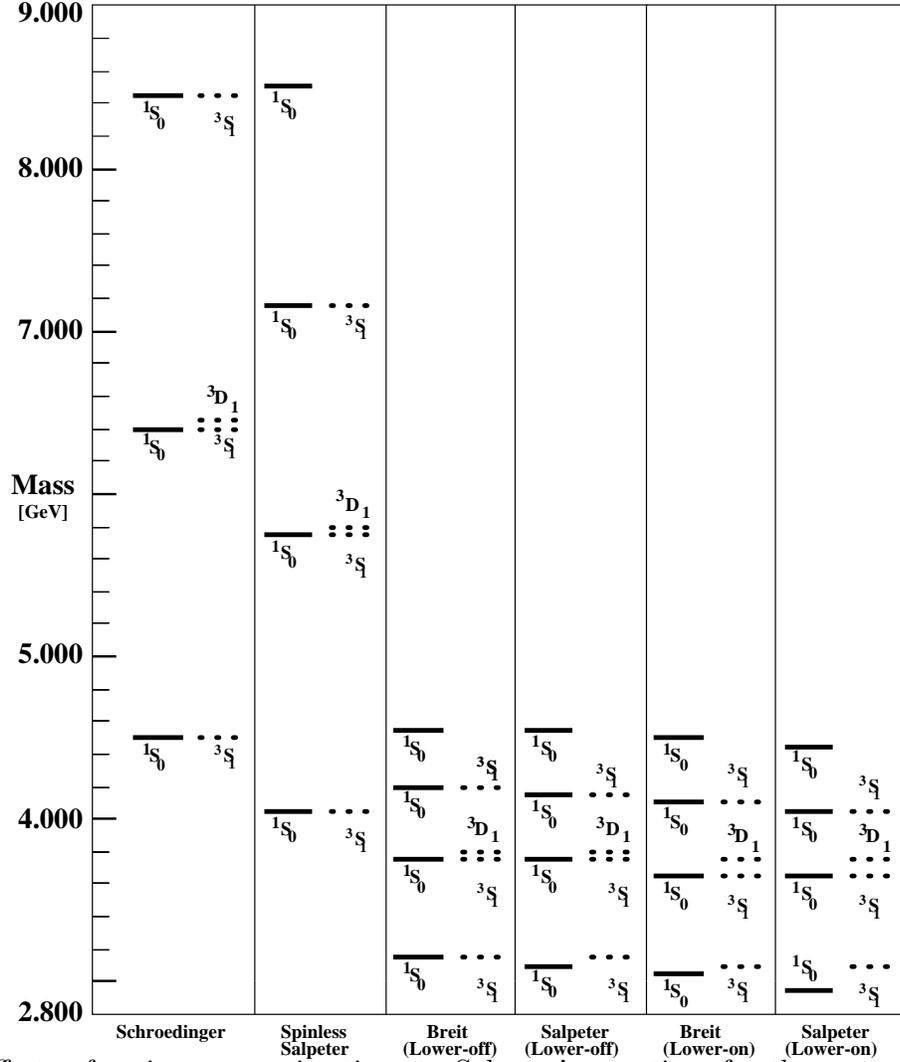}
\caption{Effects of various approximations to Salpeter's equation, for
         the parameters in Table~\protect\ref{table1chap4}, for the
         Salpeter mixed-confinement model, for charmonium.  $0^-$ and
         $1^-$ states are in each left-hand and right-hand column,
         respectively, and the spectroscopic notation quoted is the
         dominant component of the calculation.}
\label{fig10chap4}
\end{figure}

Table~\ref{table4chap4} lists the first four eigenenergies for the
pseudoscalar and vector channels, using the Salpeter-Mixed parameters
from Table~\ref{table1chap4}, to illustrate the various relativistic
effects in Salpeter's equation.  (These are also illustrated in
Figure~\ref{fig10chap4}.)

The results show a consistent decrease in the energy of a given level
as more relativistic effects are included in the calculation.  In
going from the Schr\"odinger case (nonrelativistic kinematics) to the
spinless Salpeter case (relativistic kinematics), the energy decreases
simply because the nonrelativistic kinetic energy increases quadratically
for large momenta, while the relativistic case only increases linearly;
since the potential is the same for both cases, the states are more
``bound'' in the relativistic case.  For the Breit case with no lower
components, the relativistic normalization ($\frac{E_{\bf k} + M}
{2E_{\bf k}}$) of the free Dirac spinors suppresses the overall
potential for large momenta with respect to the nonrelativistic case,
as the normalization varies from 1 in the extreme nonrelativistic limit
to $\frac{1}{2}$ in the extreme relativistic limit.  However, both
attractive and repulsive contributions to the potential are suppressed,
and the energy is still decreased relative to the nonrelativistic case.
The inclusion of Z-graphs in the Salpeter case always leads to an added
attraction, and consequently to energies reduced relative to the Breit
case.

With the introduction of the lower components, the energies are decreased
still further, for both the Breit and Salpeter cases.  In particular, the
spacelike part of the vector potential
(${\bf \gamma}_{1}\cdot{\bf \gamma}_{2}$) makes a large contribution.  The
decrease in going from Breit to Salpeter is realized from the fact that the
spacelike part connects the large component of a particle spinor to the large
component of an antiparticle spinor in $V^{+-}$, hence the contribution from
the Z-graphs is much larger than that of the direct graphs alone.

\subsubsection{Fine Structure Analysis}
One can obtain information on the spin dependence, and thus on relativistic
effects, of the effective potential for the heavy quarkonia by examining
the P-wave fine structure.  In perturbation theory (which is a good
approximation in $c\bar{c}$ and $b\bar{b}$), to ${\cal O}
\left(\frac{1}{M^2}\right)$ one can assess the relative
contributions from a Breit reduction of the potential as
\begin{equation}
  {\cal M}\left( ^{2S+1}P_{J=0,1,2} \right)
  = {\cal M}_{0} + \alpha_{SS}<{\bf S}_{1}\cdot{\bf S}_{2}>
                   + \alpha_{LS}<{\bf L}\cdot{\bf S}>
                   + \alpha_{T}<S_{12}> \;\; ,
\label{finestruc}
\end{equation}
where $\alpha_{SS}$, $\alpha_{LS}$, and $\alpha_{T}$ arise from the
spin-spin, spin-orbit, and tensor components of the potential, and
\begin{eqnarray}
   <{\bf S}_{1}\cdot{\bf S}_{2}> &=& \frac{2S(S+1) - 3}{4}
  =     \cases{+\frac{1}{4}  \,,       & for S=1,   \cr
               -\frac{3}{4}  \,,       & for S=0;   \cr  } \\
   <{\bf L}\cdot{\bf S}> &=& \frac{1}{2}\left[J(J+1) - L(L+1) - S(S+1)\right]
  =     \cases{ -2  \,,       & for $^3P_{0}$,   \cr
                -1  \,,       & for $^3P_{1}$,   \cr
                +1  \,,       & for $^3P_{2}$,   \cr
                +0  \,,       & for $^1P_{1}$;   \cr  } \\
  <S_{12}> &=& <12\left[ \left({\bf S}_{1}\cdot\hat{{\bf r}}\right)
                       \left({\bf S}_{2}\cdot\hat{{\bf r}}\right)
                       - \frac{1}{3}{\bf S}_{1}\cdot{\bf S}_{2} \right]> \\
           &=& <\frac{4}{(2L+3)(2L-1)}
                \left[ {\bf S}^2{\bf L}^2
                     - \frac{3}{2}{\bf L}\cdot{\bf S}
                     - 3({\bf L}\cdot{\bf S})^2 \right] > \\
           &=&    \cases{  -4 \,,                 & for $^3P_{0}$,   \cr
                           +2 \,,                 & for $^3P_{1}$,   \cr
                           -\frac{2}{5} \,,       & for $^3P_{2}$,   \cr
                           +0  \,,                & for $^1P_{1}$,   \cr  }
\end{eqnarray}
(the last two expressions for the tensor component applying for the
diagonal elements only), with ${\cal M}_{0}$ the unperturbed mass.
To the extent that perturbation theory is
valid for heavy-quark spectroscopy, the couplings describe fundamental
parameters of nature.  Solving the four equations [Eq.~(\ref{finestruc})]
with the four unknowns yields the data in Tables~\ref{table4achap4} and
\ref{table4bchap4}.  Perturbatively, the mass of the $^{1}P_{1}$ state
should be equal to the center of gravity (COG) of the $^{3}P_{J}$ multiplet
\begin{equation}
  {\rm COG}(^{3}P_{J})
  =\frac{1}{9}\left[ 5{\cal M}\left( ^{3}P_{2}\right)
                   + 3{\cal M}\left( ^{3}P_{1}\right)
                   + {\cal M}\left( ^{3}P_{0}\right) \right] \;\; ,
\end{equation}
with corrections up to ${\cal O}\left(\frac{1}{M^2}\right)$.
(One notes that ${\cal M}_{0}$ is equal to the COG in the
limit of a zero-range spin-spin interaction). The spin-spin
contribution is, except for the Salpeter-Mixed model, an order
of magnitude smaller than the tensor and spin-orbit contributions;
this can be understood by remembering that the spin-spin term
in the Breit reduction is a contact interaction; since the P-wave
states have no support at the origin, spin-spin effects are minimized
in this channel. (They are not zero here because relativistic
corrections in the Breit and Salpeter models regularize the
contact term.)  The $^{1}P_{1}$ state is off in all models, but
the error is about 1\% at most (0.75\% for the Salpeter-Mixed model).
For both Breit and Salpeter models, an appropriate mixture of scalar
and timelike confinement would be required for a closer match with
experiment for the $^{1}P_{1}$ $b\bar{b}$ state.

\subsection{Fits to Light Mesons}
\subsubsection{Form of the Interaction}
The interaction for the light mesons is taken to be the same form as
for the heavy mesons.  The flavor-independent OGE kernel, however,
leads to degenerate $\pi$ and $\eta$ masses; one would need to take
higher-order diagrams into account that would lead to a flavor-dependent
interaction, such as annihilation diagrams.  These, however, are
nontrivial to consider in the instantaneous framework and are not
treated in this work.  It should be noted that there are other
QCD-based candidates for flavor-dependent $q\bar{q}$ interactions that
have been computed by t'Hooft and others from instanton
effects~\cite{thooft76,shifman80}.  Such an interaction has been
employed in a study similar to this one by Resag
{\it et~al.}~\cite{munz93} for an effective description of the
light meson spectra.

\begin{figure}[tb]
\centering
\epsfig{file=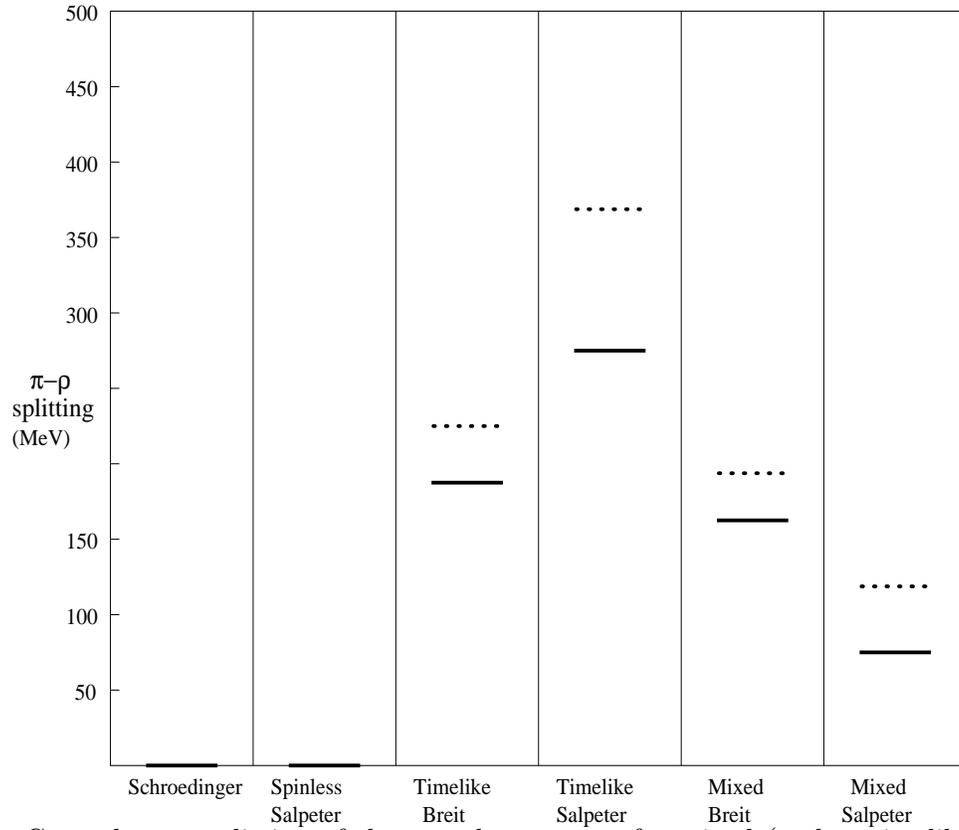,height=4.5in,width=5.0in}
\caption{Ground state splitting of the $\pi$ and $\rho$ mesons, for
         mixed (scalar+timelike) and pure timelike confinement.
         An oscillator basis with $n_{max}$=20 and $\beta$=0.3 GeV
         was used.  The model parameters were taken from
         reference~\protect\cite{spence93}, with the resultant
         splittings indicated by solid lines.  The dotted lines
         are with \mbox{$\alpha_{s}$=0.318}.}
\label{fig11chap4}
\end{figure}

One change from the heavy quarkonia is that the constant in the
confining kernel is permitted as a free parameter.  That this
is necessary is evinced by Figure~\ref{fig11chap4}, which illustrates
the \mbox{$\pi$-$\rho$} ground-state splitting, using an up mass
\mbox{$M_{u}$$\equiv$$M_{d}$=0.154 GeV}, a string tension
\mbox{$\sigma$=0.2867 ${\rm GeV}^2$}, and initially a strong
coupling \mbox{$\alpha_{s}$=0.2427 }, taken from Spence and
Vary~\cite{spence93}.  This coupling ($\alpha_{s}$) is then
increased to \mbox{$\alpha_{s}$$\approx$0.318}, the maximum
allowed value by the stability analysis in Section~\ref{secstability}.
The maximum \mbox{$\pi$-$\rho$}
splitting is less than 400 MeV, still about 230 MeV less than the
experimental value.  The simplest prescription for adjusting the
masses in order to eliminate the difference is to incorporate a
constant $c_{\sigma}$ into the confining kernel.  This ``confinement
intercept'' has been argued for on other grounds: the necessity of
regularizing the divergence which appears in treating the linear
confining kernel in momentum space leads to the appearance of a
negative constant in the potential~\cite{gromes91}.  It has also
been argued that the constant can be understood as arising from
the gluon condensate of the nonperturbative vacuum~\cite{gromes91}.
For heavy systems, its inclusion is not as important, but for light
systems (and heavy-light systems) which are affected moreso by the
long-range potential, its inclusion is necessary for even a fair
description of spin-averaged mass spectra in Schr\"odinger and
relativised Schr\"odinger (i.e., spinless Salpeter) approaches; hence,
its inclusion in the Breit and Salpeter models here is perhaps justified.

Additional complications arise from the fact that in Salpeter's
equation---and in general, any two-body quasi-potential
equation---a constant term in the confining kernel does not solely
yield an additive shift in the meson mass spectrum but provides a
dynamical contribution as well, unlike the Schr\"odinger or spinless
Salpeter cases, where one can view it as a ``negative mass'' added
to the Hamiltonian.  Even with other (repulsive) interactions present,
$c_{\sigma}$ can be increased to the point where the Salpeter
solutions exhibit RPA-type instability; the corresponding effect on
the Breit solutions is that they become unbounded from below, as with
the relativistic Coulomb problem of Section~\ref{secstability}.

\subsubsection{Data and Procedures}
As in the case of heavy quarkonia, eigenenergies of the Breit (no
coupling between positive and negative energy states) and full
Salpeter equations were solved for, expanding the Salpeter
partial-wave amplitudes in an oscillator basis of 20 states per
partial wave to insure adequate convergence of the solutions, and
with an oscillator parameter value \mbox{$\beta$=0.3 GeV} suitable
for the light mesons.  Initially, there are six free parameters in
the model: $\sigma$, $\alpha_{s}$, $M_{u}$, $M_{s}$, $x_{\sigma}$,
and $c_{\sigma}$, where $M_{s}$ is the strange quark mass.
$x_{\sigma}$ will be allowed to take only the values 0.5 and 1.0, as
before.  $\sigma$, $c_{\sigma}$, $\alpha_s$, and $M_{u}$ are fixed
by fitting to the lowest $1^{-}$ state, the two lowest $1^{+}$ states,
and the lowest $2^{+}$ state for $u\bar{u}$, minimizing the chi-squared
function
\begin{equation}
  \chi^{2}(\sigma,c_{\sigma},\alpha_{s},M_{u})
  = \sum_{i=1}^{N} \frac{(E_{ith}-E_{iexp})^2}{\sigma_{ith}^2 +
    \sigma_{iexp}^2}
\end{equation}
with the errors chosen as for the heavy mesons.  $M_{s}$ was then
obtained by taking the parameters from the fit, and adjusting it
to reproduce the $\phi$ mass.

\begin{figure}[tb]
\centering
\epsfig{file=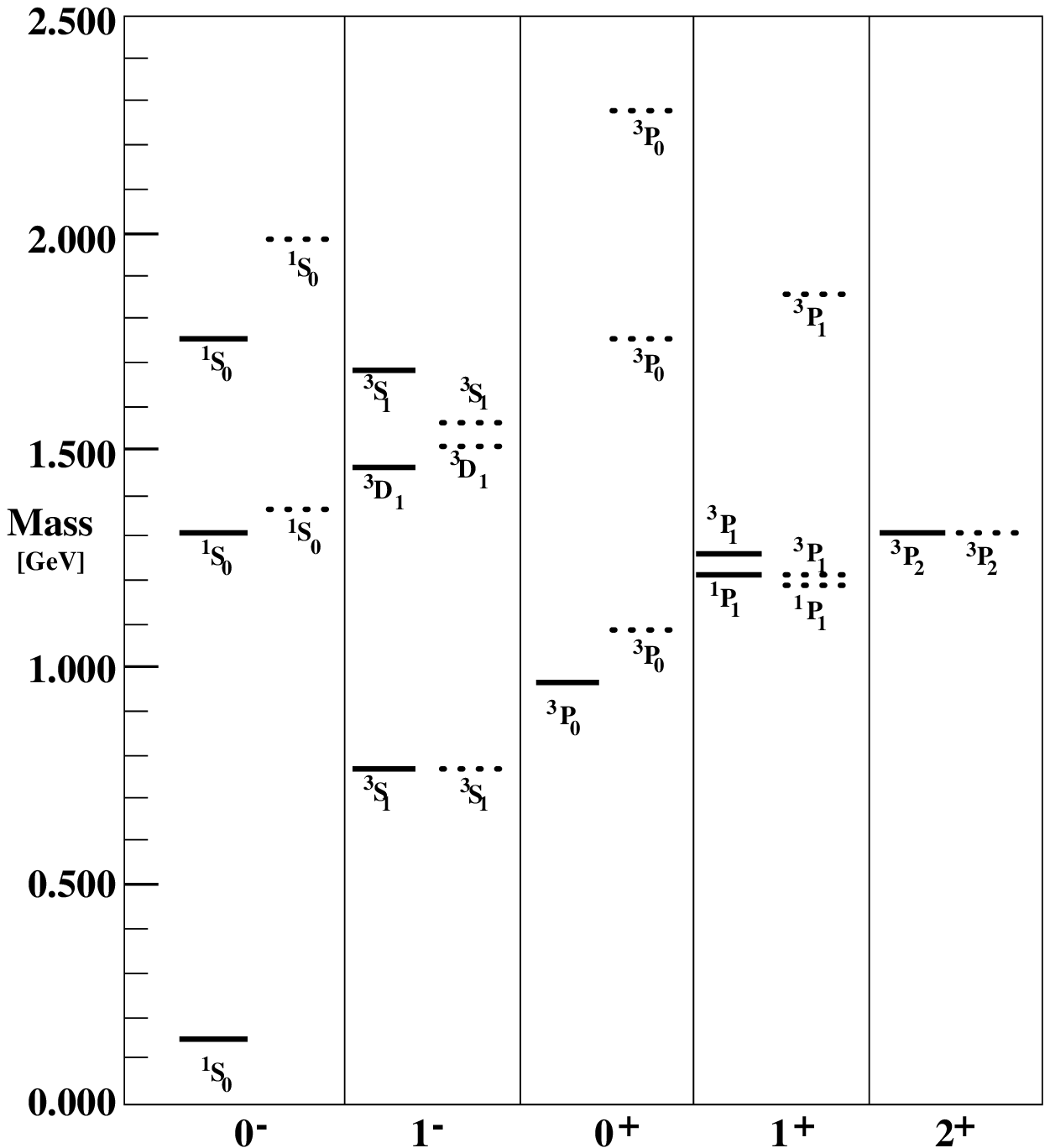}
\caption{Up mass spectrum for Salpeter, with mixed confinement
         and a vector Coulomb contribution.  The experimental
         numbers are in the left-hand column for each spin-parity.
         The spectroscopic notation for coupled states is that of
         the leading component in the calculation.}
\label{fig14chap4}
\end{figure}

\begin{figure}[tb]
\centering
\epsfig{file=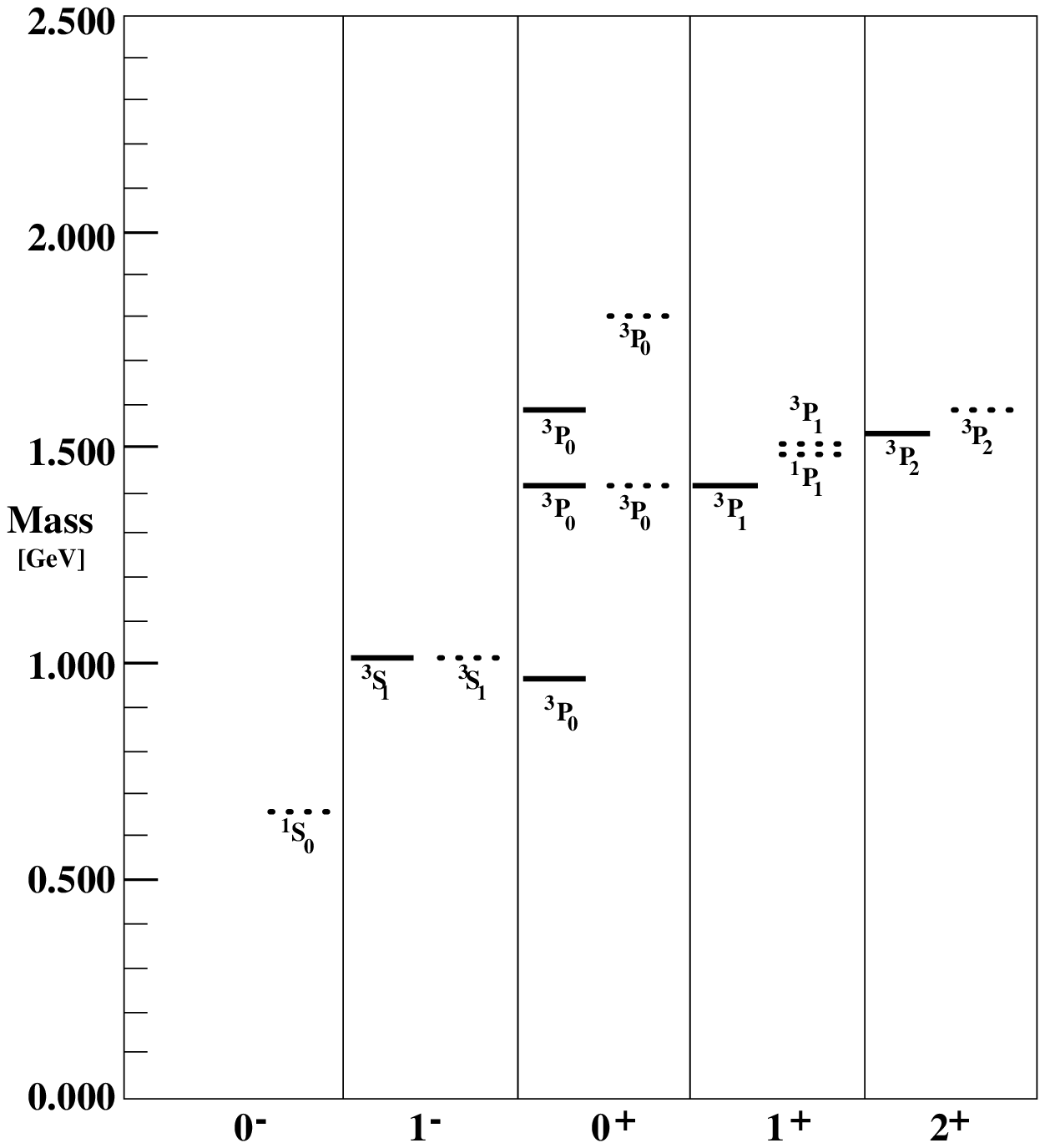,height=5.7in,width=5.0in}
\caption{Strange quarkonium mass spectrum for Salpeter, with mixed
         confinement and a vector Coulomb contribution.  The
         experimental numbers are in the left-hand column for each
         spin-parity.  The spectroscopic notation for coupled states
         is that of the leading component in the calculation.}
\label{fig18chap4}
\end{figure}

The minimization results for the four parameters for the four models
(Breit-Timelike, Breit-Mixed, Salpeter-Timelike, and Salpeter-Mixed)
are summarized in Table~\ref{table5chap4}, while the light and strange
quarkonia spectra calculated for each model are summarized in
Table~\ref{table6chap4} and Table~\ref{table7chap4}; graphical
depictions of the spectra for each $J^{\pi}$ channel are shown for
the Salpeter-Mixed model in Figures~\ref{fig14chap4}~and~\ref{fig18chap4}.
$J^{\pi}$ states are named by their main $^{2S+1}L_{J}$ components.

The Salpeter-Mixed model is the best model in this case.  The problem
with it, however, and with the Salpeter-Timelike model, is that in
order to fit the $\rho$, $c_{\sigma}$ had to be increased to the
point where the $\pi$ became unstable.  The constituent mass values fall
within accepted ranges for the Salpeter-Mixed case and the confinement
slope is larger than the empirical value
\mbox{$\sigma$$\approx$0.2GeV$^2$} obtained from spectroscopy.
However, it is still in agreement with the lattice result
\mbox{$\sigma$=${0.33}_{-0.23}^{+0.82}$GeV$^2$}\cite{gromes91}; although
this last comparison is not very significant because of the large error
bars.  The confinement offset is comparable to that obtained from the
prescription \mbox{$c_{\sigma}$$\approx$$-2\sqrt{\sigma}$}\cite{gromes91}.
The Breit cases fit the $\rho$, but cannot reproduce the $\pi$ at all.
That some difficulty should be encountered in describing the pion in
these models should not be unexpected.  The mass of the $\pi$ is
commonly explained in the framework of broken chiral symmetry, where it
corresponds to an almost massless Goldstone boson; such models
incorporating chiral symmetry have been investigated by Gross and
Milana~\cite{gross94}.

In this case, the necessity for the coupling between positive and
negative energy states for the light mesons is well illustrated;
the Z-graphs provide an additional attraction that may be necessary
in describing the $\pi$ as a deeply bound state of a quark and an
antiquark (although in the present model the attraction is too strong
in this channel).  This need for the $V^{+-}$ component in the Salpeter
equation agrees with the results of Gara {\it et~al.}~\cite{gara89} as
well, albeit for different reasons.  The $s\bar{s}$ states are
well-described, with the exception of the $0^{+}$ states.  It is known,
however, that these scalar states can not be represented as simple
$q\bar{q}$ states\cite{weinstein90}.  Note that in this case the
strange-quark mass was adjusted to reproduce the $\phi$; the other
states are predictions of the model.

\section{Conclusions}
\label{secconcl}

We have used Salpeter's equation to study the light- and heavy-meson
spectra. This study was preceded by a stability analysis of Salpeter's
equation that proved essential for placing limits on the form of the
instantaneous kernel. We stress that because of the RPA structure of
Salpeter's equation a stability analysis must always be
performed---regardless of the form of the interaction kernel.

The two main results that emerged from the stability analysis are:
1) the Lorentz character of the confining kernel must be timelike
or a mixture of scalar and timelike forms, contrary to the usual
assumption of pure scalar confinement, and 2) an upper limit of
$\alpha_{s}={1/\pi}$ was set on the strong coupling constant
used in the OGE kernel. This value, and the corresponding value
for the ground-state energy, were obtained from a variational
solution to the semirelativistic Coulomb problem. To our knowledge
this is the best estimate presented to date. Having placed limits
on the interaction kernel we proceeded to carry out a detailed study
of the heavy and light quarkonia.

Static properties of the heavy and light quarkonia within Salpeter's
framework have been examined using a generalization of the Cornell
potential. For the heavy quarkonia the relativistic corrections
coming into play in the various models were examined. These
models included Salpeter and Breit approximations having, either,
a timelike or a mixture of scalar and timelike Lorentz structures
for the confining potential. Recall that the Breit approximation
is obtained by setting $V^{+-}$ to zero. Meson masses were
adequately described in all the models, with the best results obtained
using the Breit model with timelike confinement. However, a perturbative
study of spin-dependent effects (valid for the heavy quarkonia)
reveals that the fine structure (P-wave splittings) and hyperfine
structure (\mbox{$^{3}S_{1}$-$^{3}D_{1}$} splitting) cannot be
simultaneously described in any of the models by simply varying
the mixing of scalar and timelike confinement. For the light
quarkonia, the mass spectra, except for the pion, are best
described by the Salpeter model with mixed scalar-timelike
confinement. However, none of the models were able to describe
the pion, or equivalently, the \mbox{$\pi$-$\rho$} splitting.
For example, in the Breit model the \mbox{$\pi$-$\rho$} splitting
is a ``mere'' 180~MeV. This difference can be pushed up to about
400~MeV in the Salpeter model at the critical coupling; still
this value is substantially smaller than the experimental splitting
of 630~MeV. The additional attraction needed to describe the pion
would appear to rule out using the Breit models for a description
of static meson properties. Whether or not this is also sufficient
to rule out the use of Salpeter's equation is not clear at this point.
Overall, all of the features of meson spectroscopy could not be
simultaneously satisfied using the relatively simple kernel employed
here. At the very least, a more sophisticated phenomenology is required,
especially for the light mesons and in particular for the pion.  It is
likely that some form of chirally-invariant model will be
needed~\cite{gross94}. It seems clear, however, that keeping the
couplings between positive- and negative-energy states is necessary
for any realistic description of at least the light spectra, and
certainly for a combined heavy-light analysis. Moreover, it should
also be clear that regardless of the form of the kernel, the stability
analysis used here must be employed in any study that has Salpeter's
equation as the underlying dynamical framework.

\acknowledgments

We thank D.~Robson for many helpful discussions.
This research was supported by the Florida State University
Supercomputer Computations Research Institute and the U.S.
Department of Energy contracts DE-FC05-85ER250000,
DE-FG05-92ER40750, and DE-FG05-86ER40273.
\appendix

\section{Extraction of the coefficients to ${\cal O}(\epsilon)$ in the
         variational energy}

For a variational wave function of the forms given in Eq.~(\ref{eq.expansion})
and Eq.~(\ref{eq.pexpansion}), one wishes to calculate the matrix elements of
the kinetic energy operator and potential operator as $\epsilon \rightarrow 0$.
These matrix elements diverge logarithmically for large momenta and small radii
in their respective integrands; however, one can examine the behavior of the
variational energy for small $\epsilon$, and determine the critical value
of the Coulomb coupling at which the system becomes unbounded.  In order
to compute the variational energy for small values of $\epsilon$ we expand
$\langle T \rangle$ and $\langle V \rangle$ as Laurent series\cite{arfken85}
in $\epsilon$:
\begin{eqnarray}
  {\langle T \rangle \over M} &=& {t_{-1}\over\epsilon} + t_0 + t_1 \epsilon +
   \cdots \\
  {\langle V \rangle \over M} &=& {v_{-1}\over\epsilon} + v_0 + v_1 \epsilon +
   \cdots~.
\end{eqnarray}
One wants the leading coefficients $t_{-1},\ v_{-1}$ and the zeroth-order ones
$t_{0},\ v_{0}$ as well.  By minimizing $h_0 = t_0 + v_0$ we could find the
variational energy in the $\alpha=\alpha_c$ limit.  First we calculate the
expectation value of $H$ using only one term in the polynomial expansion in
Eq.~(\ref{eq.expansion}), i.e., we set $c_1=c_2=\cdots=0$.
Starting with the potential, one has
\begin{equation}
  {\langle V \rangle \over M}
   = -2\alpha \left[ {\gamma\Gamma(2\epsilon)\over
                     M\Gamma(1+2\epsilon)} \right]
   = -2\alpha \left[ {a\Gamma(2\epsilon) \over
     (2\epsilon)\Gamma(2\epsilon)} \right]
   = ( -\alpha a ) {1 \over \epsilon}~.
  \label{eq.v_expt}
\end{equation}
where $\Gamma(z)$ is the gamma function~\cite{arfken85,gradshteyn80}.
Note that Eq.~(\ref{eq.v_expt}) is an exact result; there are no terms
of higher order in $\epsilon$ in the series.  The kinetic piece
requires a little more care; one first rewrites the integrand using
a standard trigonometric identity:
\begin{eqnarray}
  {\langle T \rangle \over M}
   &=& \int_0^{\infty}4\pi k^2 dk {2\sqrt{k^2+M^2}\over M}
       {1\over 4\pi} \left\{
   {1\over k} \sqrt{2\over\pi} N_R {\Gamma(1+\epsilon) \over
   (k^2+\gamma^2)^{(1+\epsilon)/2} } \sin \left[ (1+\epsilon)
   \tan^{-1}\left({k\over\gamma} \right) \right] \right\}^2 \nonumber \\
  &=& {4 (2{\gamma\over M})^{1+2\epsilon} \left[\Gamma(1+\epsilon)\right]^2
   \over \pi \Gamma(1+2\epsilon)} \int_{0}^{\infty} {dk\over M}
   {\sqrt{\left({k\over M}\right)^2 + 1} \over \left( {k^2+\gamma^2\over M^2}
   \right)^{1+\epsilon} } \left\{ 1 -
   \cos^{2}\left[ (1+\epsilon) \tan^{-1}\left({k\over\gamma}\right) \right]
   \right \}~.
  \label{eq.t_expt}
\end{eqnarray}
The second integral in Eq.~(\ref{eq.t_expt}) is convergent, so there is no
ambiguity in setting $\epsilon = 0$ explicitly.  The first integral contains a
logarithmic divergence for large $k$; one adds and subtracts this divergence to
obtain (with $a\equiv\gamma/M$ and $k/M \rightarrow k$)
\begin{eqnarray}
  {\langle T \rangle \over M}  &=& {\langle T \rangle_C \over M} +
                                   {\langle T \rangle_D \over M} \\
  {\langle T \rangle_C \over M} &=&
   {4 (2a)^{1+2\epsilon} \left[\Gamma(1+\epsilon)\right]^2
   \over \pi \Gamma(1+2\epsilon)} \int_{0}^{\infty} dk \left\{
   {\sqrt{k^2+1}-k \over (k^2+a^2)^{1+\epsilon} } -
   {\sqrt{k^2+1}   \over (k^2+a^2)^{1+\epsilon} }
   \cos^{2}\left[(1+\epsilon)\tan^{-1}\left({k\over a}\right)\right] \right\}\\
  {\langle T \rangle_D \over M} &=&
   {4 (2a)^{1+2\epsilon} \left[\Gamma(1+\epsilon)\right]^2
   \over \pi \Gamma(1+2\epsilon)} \int_{0}^{\infty} dk
   {k\over (k^2+a^2)^{1+\epsilon} }
\end{eqnarray}
$\langle T \rangle_D$ is logarithmically divergent for large momenta, for
$\epsilon=0$, so one evaluates the integral for $\epsilon$ nonzero, then
expands about $\epsilon=0$ up to ${\cal O}(\epsilon)$.  One then has
\begin{eqnarray}
  {\langle T \rangle_D \over M} &=&
   {4 (2a)^{1+2\epsilon} \left[\Gamma(1+\epsilon)\right]^2
   \over \pi \Gamma(1+2\epsilon)}
   {1 \over 2 \epsilon a^{2\epsilon} }
  \label{eq.t_div} \\
  &=& \left( {4a\over\pi} \right) {1\over\epsilon}
   + {8a\over\pi} \ln 2 + {\cal O}(\epsilon)~.
  \label{eq.t_divexpd}
\end{eqnarray}
One should note that in going from Eq.~(\ref{eq.t_div}) to
Eq.~(\ref{eq.t_divexpd}), one must be careful to include
{\it all factors} in constructing the series expansion.

The remaining integrals in $\langle T \rangle_C$ are convergent,
and one can set $\epsilon=0$ explicitly.  Using
\begin{displaymath}
  \cos\left[ \tan^{-1}\left({k\over a}\right) \right] =
   {a \over \sqrt{k^2+a^2}}
\end{displaymath}
one has
\begin{eqnarray}
  {\langle T \rangle_C \over M} &=& {8a\over\pi} \int_{0}^{\infty} dk \left[
   {\sqrt{k^2+1}-k \over k^2+a^2 } -
   {\sqrt{k^2+1} a^2 \over (k^2+a^2)^2 } \right] \\
  &=& {8a\over\pi} \int_{0}^{\infty} dk
   { k^2 \sqrt{k^2+1} - k \sqrt{k^2+a^2} \over (k^2+a^2)^2 } \\
  &=& {8a\over\pi} \tilde{T}_C(a)~.
\end{eqnarray}
Then, in the limit that $\epsilon\rightarrow 0$, $\langle H \rangle / M$ can be
written as
\begin{equation}
  {\langle H \rangle \over M} = \left[ \left({4\over\pi}-\alpha\right)a \right]
   {1\over\epsilon} + {8a\over\pi} \ln 2 + {8a\over\pi} \tilde{T}_C(a)
   + {\cal O}(\epsilon)~,
  \label{eq.h_expt}
\end{equation}
where $\tilde{T}_C(a)$ is given in closed form for $a > 0$ by
\begin{equation}
  \tilde{T}_C(a) = \cases{
   \ln 2 + \ln a - {1\over2}
         + {1-2a^2 \over 2a\sqrt{1-a^2}}
         \tan^{-1} \left( {\sqrt{1-a^2}\over a} \right) \;\; & $a<1 \;\;$ ; \cr
   \ln 2 - 1                                                 & $a=1 \;\;$ ; \cr
   \ln 2 + \ln a - {1\over2}
         + {1-2a^2 \over 2a\sqrt{a^2-1}}
         \left[ {1\over2} \ln \left( {a+\sqrt{a^2-1} \over a-\sqrt{a^2 - 1}}
         \right) \right]                                     & $a>1 \;\;$ . \cr
   }
\end{equation}
By demanding that the coefficient of the singular term $h_{-1}$ in the series
vanishes we can extract the critical coupling $\alpha_c=4/\pi$.  Furthermore,
by minimizing the zeroth-order coefficient $h_0$ we can derive the variational
energy ($E/M=0.968583$) and the optimal parameter ($a=0.7926$).

%
%

%
%
%
%

%
\mediumtext
\begin{table}[tb]
\centering
  \caption{Minimization of the variational energy
           with respect to the parameters $\epsilon$ and $a$
           for fixed $\alpha$, as $\alpha$ approaches $\alpha_c$.}
\smallskip
\begin{tabular}{dddd}
   $\alpha$ & $\frac{{<H>}}{M}$ & $a$ & $\epsilon$ \\ \hline
   0.100   &   1.9975   &   0.0495   & 0.9861   \\
   0.200   &   1.9899   &   0.0988   & 0.9660   \\
   0.400   &   1.9582   &   0.1956   & 0.8977   \\
   0.600   &   1.9007   &   0.2930   & 0.8041   \\
   0.800   &   1.8081   &   0.3944   & 0.6825   \\
   1.000   &   1.6583   &   0.5071   & 0.5222   \\
   1.200   &   1.3639   &   0.6563   & 0.2703   \\
   1.220   &   1.3106   &   0.6778   & 0.2302   \\
   1.240   &   1.2435   &   0.7033   & 0.1817   \\
   1.260   &   1.1461   &   0.7374   & 0.1144   \\
   1.270   &   1.0581   &   0.7657   & 0.05646  \\
   1.273   &   0.9933   &   0.7854   & 0.01533  \\
   1.2731  &   0.9874   &   0.7871   & 0.01170  \\
   1.2732  &   0.9786   &   0.7897   & 0.006226 \\
\end{tabular}
\label{table4chap3}
\end{table}
%
\mediumtext
\begin{table}
\centering
\caption{Definitions of the various models used in the calculations;
         ``Mixed'' and ``Timelike'' refer to the Lorentz structure
         of the confining kernel.  The OGE kernel is always of vector
         type.}
\smallskip
\begin{tabular}{ccc}
          & Mixed   &   Timelike \\ \hline
   Breit  & [$V^{+-}$=0, $x_{\sigma}$=0.5] & [$V^{+-}$=0, $x_{\sigma}$=1] \\
 Salpeter & [$V^{+-}$$\neq$0, $x_{\sigma}$=0.5] & [$V^{+-}$$\neq$0,
            $x_{\sigma}$=1] \\
\end{tabular}
\label{table1achap4}
\end{table}
%
%
\mediumtext
\begin{table}
\centering
\caption{Heavy meson parameters for Breit and Salpeter models
         with mixed (scalar+timelike) and timelike confinement,
         plus an instantaneous vector OGE contribution.
         An oscillator basis with \mbox{$n_{max}$=20} and
         \mbox{$\beta$=0.6 GeV} was employed for the parameter
         optimization.}
\smallskip
\begin{tabular}{ccccc}
  Parameters  &  $\stackrel{\textstyle {\rm Breit}}{\rm Mixed}$
              &  $\stackrel{\textstyle {\rm Breit}}{\rm Timelike}$
              &  $\stackrel{\textstyle {\rm Salpeter}}{\rm Mixed}$
              &  $\stackrel{\textstyle {\rm Salpeter}}{\rm Timelike}$
              \\ \hline
  $M_{c}$ [GeV]            &  1.168   &  1.379  &  1.251  &  1.126  \\
  $M_{b}$ [GeV]            &  4.573   &  4.781  &  4.623  &  4.561  \\
  $\sigma$ [${\rm GeV}^2$] &  0.2991  &  0.1937 &  0.2570 &  0.2743 \\
  ${\alpha}_{s}$           &  0.2678  &  0.4875 &  0.2872 &  0.2165 \\
\end{tabular}
\label{table1chap4}
\end{table}
%
\mediumtext
\begin{table}
\centering
\caption{Charm quarkonia masses in GeV for Breit and Salpeter models
         with mixed (scalar+timelike) and timelike confinement,
         plus an instantaneous vector OGE contribution, for the
         parameters as in Table~\protect\ref{table1chap4}.
         The calculated states are aligned with the observed states
         according to their spin parity, starting from the lowest mass
         values.  An asterisk on an observed value indicates a state
         employed in the fits.  An oscillator basis with $n_{max}$=20
         and $\beta$=0.6 GeV was employed for all states.}
\smallskip
\begin{tabular}{ccc@{}dcccc}
  Meson   &  ${J}^{\pi}$ & $^{{2S+1}}L_{J}$  & $~~~M_{expt}$ &
 $\stackrel{\textstyle {\rm Breit}}{\rm Mixed}$ &
 $\stackrel{\textstyle {\rm Breit}}{\rm Timelike}$ &
 $\stackrel{\textstyle {\rm Salpeter}}{\rm Mixed }$ &
 $\stackrel{\textstyle {\rm Salpeter}}{\rm Timelike}$
\\ \hline
 $\eta_c$  &  $0^-$ & $^{1}S_{0}$ & 2.979$^{*}$    &
              2.984 &  2.977 &  2.958 &  2.968 \\
 $J/\psi$  &  $1^-$ & $^{3}S_{1}$ & 3.097$^{*}$    &
              3.044 &  3.111 &  3.099 &  3.056 \\
 $\chi_{c0}$  &  $0^+$ & $^{3}P_{0}$ & 3.415             &
              3.360 &  3.326 &  3.372 &  3.338 \\
 $\chi_{c1}$  &  $1^+$ & $^{3}P_{1}$ & 3.511             &
              3.421 &  3.468 &  3.412 &  3.424 \\
 $h_{c}$   &  $1^+$ & $^{1}P_{1}$ & (3.526)              &
              3.440 &  3.514 &  3.468 &  3.455 \\
 $\chi_{c2}$  &  $2^+$ & $^{3}P_{2}$ & 3.556             &
              3.467 &  3.562 &  3.499 &  3.513 \\
 $\eta_c$  &  $0^-$ & $^{1}S_{0}$ & (3.594)$^{*}$  &
              3.645 &  3.597 &  3.622 &  3.647 \\
 $\psi$    &  $1^-$ & $^{3}S_{1}$ & 3.685$^{*}$    &
              3.688 &  3.675 &  3.693 &  3.705 \\
 $\psi$    &  $1^-$ & $^{3}D_{1}$ & 3.770$^{*}$    &
              3.726 &  3.755 &  3.739 &  3.732 \\
 $\psi$    &  $1^-$ & $^{3}S_{1}$ & 4.040$^{*}$    &
              4.167 &  4.099 &  4.134 &  4.218 \\
 $\psi$    &  $1^-$ & $^{3}D_{1}$ & 4.159$^{*}$    &
              4.192 &  4.152 &  4.162 &  4.234 \\
 $\psi$    &  $1^-$ & $^{3}S_{1}$ &                      &
              4.566 &  4.459 &  4.499 &  4.659 \\
 $\psi$    &  $1^-$ & $^{3}D_{1}$ & 4.415              &
              4.583 &  4.499 &  4.519 &  4.669 \\
\end{tabular}
\label{table2chap4}
\end{table}
\mediumtext
\begin{table}
\centering
\caption{Beauty quarkonia masses in GeV for Breit and Salpeter models
         with mixed (scalar+timelike) and timelike confinement,
         plus a vector OGE (instantaneous Coulomb) contribution, for the
         parameters as in Table~\protect\ref{table1chap4}.
         The calculated states are aligned with the observed states
         according to their spin parity, starting from the lowest mass
         values.  An asterisk on an observed value indicates a state
         employed in the fits.  An oscillator basis with $n_{max}$=20
         and $\beta$=0.6 GeV was employed for all states.}
\smallskip
\begin{tabular}{ccc@{}ddddd}
  Meson   &  ${J}^{\pi}$ & $^{{2S+1}}L_{J}$  &  $~~~M_{expt}$ &
 $\stackrel{\textstyle {\rm Breit}}{\rm Mixed}$ &
 $\stackrel{\textstyle {\rm Breit}}{\rm Timelike}$ &
 $\stackrel{\textstyle {\rm Salpeter}}{\rm Mixed }$ &
 $\stackrel{\textstyle {\rm Salpeter}}{\rm Timelike}$
\\ \hline
 $\eta_{b}$&  $0^-$ & $^{1}S_{0}$ &                    &
  9.440 &  9.377 &  9.373 &  9.432 \\
 $\Upsilon$&  $1^-$ & $^{3}S_{1}$ & 9.460$^{*}$  &
  9.468 &  9.459 &  9.485 &  9.488 \\
 $\chi_{b0}$&  $0^+$ & $^{3}P_{0}$ & 9.860              &
  9.821 &  9.853 &  9.825 &  9.807 \\
 $\chi_{b1}$&  $1^+$ & $^{3}P_{1}$ & 9.892              &
  9.843 &  9.905 &  9.850 &  9.827 \\
 $\chi_{b1}$&  $1^+$ & $^{1}P_{1}$ &                    &
  9.850 &  9.921 &  9.841 &  9.826 \\
 $\chi_{b2}$&$2^+$ & $^{3}P_{2}$ & 9.913               &
  9.858 &  9.940 &  9.865 &  9.845 \\
 $\eta_{b}$&  $0^-$ & $^{1}S_{0}$ &                    &
  9.995 &  9.992 &  9.949 &  9.962 \\
 $\Upsilon$&  $1^-$ & $^{3}S_{1}$ & 10.023$^{*}$ & 10.013 &
  10.023 &  9.999 &  9.994 \\
 $\Upsilon$&  $1^-$ & $^{3}D_{1}$ &                    &
  10.112 & 10.164 & 10.100 & 10.077 \\
 $\chi_{b0}$&$0^+$ & $^{3}P_{0}$ &10.232               &
  10.245 & 10.230 & 10.214 & 10.211 \\
 $\chi_{b1}$&$1^+$ & $^{3}P_{1}$ &10.255               &
  10.263 & 10.266 & 10.236 & 10.229 \\
 $\chi_{b2}$&$2^+$ & $^{3}P_{2}$ &10.268               &
  10.277 & 10.293 & 10.249 & 10.244 \\
 $\eta_{b}$&  $0^-$ & $^{1}S_{0}$ &                    &
  10.397 & 10.350 & 10.332 & 10.352 \\
 $\Upsilon$&  $1^-$ & $^{3}S_{1}$ &10.355              &
  10.412 & 10.371 & 10.367 & 10.375 \\
 $\Upsilon$&  $1^-$ & $^{3}D_{1}$ &                    &
  10.477 & 10.462 & 10.434 & 10.432 \\
 $\Upsilon$&  $1^-$ & $^{3}S_{1}$ &10.580              &
  10.748 & 10.655 & 10.675 & 10.702 \\
\end{tabular}
\label{table3chap4}
\end{table}
\mediumtext
\begin{table}
\centering
\caption{Various relativistic effects in Salpeter's equation
         displayed in the pseudoscalar and vector channels
         for charmonium, using the parameters from the
         Salpeter-mixed model as in Table~\protect\ref{table1chap4}.
         $S$ and $D$ label the dominant $L$-wave component in the
         calculated energy.  {\it Lower on} and {\it Lower off} refer
         to the lower components of Dirac spinors being present or
         not, respectively, in the calculation.}
\smallskip
\begin{tabular}{c@{\hspace{-5pt}}d@{\hspace{-5pt}}d@{\hspace{-5pt}
                }d@{\hspace{-5pt}}d@{\hspace{-5pt}}d@{\hspace{-5pt}}d}
 ${J}^{\pi}$ &  ~~~~~Schr\"odinger &
 $~~~~~~~~\stackrel{\textstyle {\rm Spinless}}{\rm Salpeter}$ &
 $~~~~~~~~\stackrel{\textstyle {\rm Breit}}{\mbox{Lower-off}}$ &
 $~~~~~~~~\stackrel{\textstyle {\rm Salpeter}}{\mbox{Lower-off}}$ &
 $~~~~~~~~\stackrel{\textstyle {\rm Breit}}{\mbox{Lower-on}}$ &
 $~~~~~~~~\stackrel{\textstyle {\rm Salpeter}}{\mbox{Lower-on}}$
\\ \hline
$0^{-}$  &  4.302 (S)  &  4.140 (S)   &   3.154 (S) &
  3.104 (S) & 3.049 (S) & 2.958 (S) \\
              & 6.407 (S)  &  5.774 (S)    &
  3.761 (S) & 3.740 (S)  & 3.659 (S) & 3.623 (S) \\
              & 8.464 (S)  &  7.193 (S)    &
  4.200 (S) & 4.187 (S)  & 4.108 (S) & 4.084 (S) \\
              &10.499 (S)  & 8.480 (S)    &
  4.562 (S) & 4.552 (S)  & 4.481 (S) & 4.461 (S) \\
\hline
$1^{-}$  & 4.302 (S)  & 4.140 (S)   &   3.154 (S) &
  3.149 (S) & 3.107 (S) & 3.099 (S) \\
              & 6.407 (S)  & 5.774 (S)   &   3.761 (S)  &
  3.758 (S) & 3.699 (S) & 3.693 (S) \\
              & 6.442 (D)  & 5.880 (D)   &  3.817 (D)  &
  3.816 (D)  & 3.747 (D) & 3.739 (D) \\
              & 8.464 (S)  & 7.193 (S)   &  4.200 (S)   &
  4.198 (S)  & 4.141 (S)  & 4.134 (S) \\
\end{tabular}
\label{table4chap4}
\end{table}
\mediumtext
\begin{table}
\centering
\caption{Spin-dependent parameters for spin-spin, spin-orbit,
         and tensor contributions to the effective potential
         for charmonium.  ``COG'' refers to the center of gravity
         of a given multiplet. All values listed are in units of
         [GeV].}
\smallskip
\begin{tabular}{cddddd}
  Parameter   & Experiment &
 $\stackrel{\textstyle {\rm Breit}}{\rm Mixed}$ &
 $\stackrel{\textstyle {\rm Breit}}{\rm Timelike}$ &
 $\stackrel{\textstyle {\rm Salpeter}}{\rm Mixed }$ &
 $\stackrel{\textstyle {\rm Salpeter}}{\rm Timelike}$
\\ \hline
 COG($^{3}P_{J}$) &  3.525 &  3.440   &  3.504  &  3.456  &  3.464 \\
 $^{1}P_{1}$      &  3.526 &  3.440   &  3.514  &  3.468  &  3.455 \\
 ${\cal M}_{0}$   &  3.526 &  3.440   &  3.507  &  3.459  &  3.462 \\
 $\alpha_{SS}$    & -0.001 & -0.0002  & -0.010  & -0.012  &  0.009 \\
 $\alpha_{LS}$    &  0.035 &  0.029   &  0.063  &  0.043  &  0.051 \\
 $\alpha_{T}$     &  0.010 &  0.005   &  0.013  & -0.001  &  0.006 \\
\end{tabular}
\label{table4achap4}
\end{table}
\mediumtext
\begin{table}
\centering
\caption{Spin-dependent parameters for spin-spin, spin-orbit,
         and tensor contributions to the effective potential
         for beauty quarkonium.  ``COG'' refers to the center
         of gravity of a given multiplet. The given $^{1}P_{1}$
         experimental mass is actually the calculated COG.  All
         values listed are in units of [GeV].}
\smallskip
\begin{tabular}{cddddd}
  Parameter   & Experiment &
 $\stackrel{\textstyle {\rm Breit}}{\rm Mixed}$ &
 $\stackrel{\textstyle {\rm Breit}}{\rm Timelike}$ &
 $\stackrel{\textstyle {\rm Salpeter}}{\rm Mixed }$ &
 $\stackrel{\textstyle {\rm Salpeter}}{\rm Timelike}$
\\ \hline
 COG($^{3}P_{J}$) &  9.900   &  9.849   &  9.919  &  9.852  &  9.833 \\
 $^{1}P_{1}$      &  9.900   &  9.850   &  9.921  &  9.841  &  9.826 \\
 ${\cal M}_{0}$   &  9.900   &  9.849   &  9.919  &  9.852  &  9.833 \\
 $\alpha_{SS}$    &  0.0001  & -0.001   & -0.002  &  0.015  &  0.009 \\
 $\alpha_{LS}$    &  0.014   &  0.010   &  0.023  &  0.010  &  0.011 \\
 $\alpha_{T}$     &  0.003   &  0.002   &  0.005  &  0.002  &  0.002 \\
\end{tabular}
\label{table4bchap4}
\end{table}
%
%
\mediumtext
\begin{table}
\centering
\caption{Light meson parameters for Breit and Salpeter models
         with mixed (scalar+timelike) and timelike confinement,
         plus an instantaneous OGE contribution.
         An oscillator basis with $n_{max}$=20 and $\beta$=0.3 GeV
         was employed for the data fitting.}
\smallskip
\begin{tabular}{ccccc}
  Parameter   &  $\stackrel{\textstyle {\rm Breit}}{\rm Mixed}$
              &  $\stackrel{\textstyle {\rm Breit}}{\rm Timelike}$
              &  $\stackrel{\textstyle {\rm Salpeter}}{\rm Mixed}$
              &  $\stackrel{\textstyle {\rm Salpeter}}{\rm Timelike}$
              \\ \hline
  $M_{u}$ [GeV]            &  0.2862  &  0.3393 &  0.3229 &  0.4196 \\
  $M_{s}$ [GeV]            &  0.5500  &  0.5720 &  0.5610 &  0.6240 \\
  $\sigma$ ${\rm [GeV]}^2$ &  0.3841  &  0.2576 &  0.3744 &  0.2574 \\
  $c_{\sigma}$ [GeV]       & -1.448   & -1.089  & -1.427  & -1.157  \\
  ${\alpha}_{s}$           &  0.2919  &  0.3064 &  0.2690 &  0.2690 \\
\end{tabular}
\label{table5chap4}
\end{table}
%
\mediumtext
\begin{table}
\centering
\caption{Light quarkonia masses in GeV for Breit and Salpeter models
         for the parameters as in Table~\protect\ref{table5chap4}.
         The calculated states are aligned with the observed states
         according to their spin parity. An asterisk on an observed value
         indicates a state employed in fitting.  An ``I'' indicates imaginary
         eigenvalues.}
\smallskip
\begin{tabular}{ccc@{}dcccc}
  Meson   &  ${J}^{\pi}$ & $^{{2S+1}}L_{J}$  &  $~~~M_{expt}$ &
 $\stackrel{\textstyle {\rm Breit}}{\rm Mixed}$ &
 $\stackrel{\textstyle {\rm Breit}}{\rm Timelike}$ &
 $\stackrel{\textstyle {\rm Salpeter}}{\rm Mixed }$ &
 $\stackrel{\textstyle {\rm Salpeter}}{\rm Timelike}$
\\ \hline
$\pi$        &  $0^{-}$ & $^{1}S_{0}$ &  0.140  &
  0.627  &  0.642  & I & I \\
$\rho$       &  $1^{-}$ & $^{3}S_{1}$ &  0.768$^{*}$   &
  0.771  &  0.770  & 0.768 & 0.769  \\
$a_{0}$      &  $0^{+}$ & $^{3}P_{0}$ &  0.983  &  1.066  &
  0.839  & 1.014 & 0.787               \\
$b_{1}$      &  $1^{+}$ & $^{1}P_{1}$ &  1.232$^{*}$   &
  1.203  &  1.153  & 1.195 & 1.148  \\
$a{_1}$      &  $1^{+}$ & $^{3}P_{1}$ &  1.260$^{*}$ &
  1.169   & 1.033   & 1.205 & 1.081 \\
$\pi^\prime$ &  $0^{-}$ & $^{1}S_{0}$ &  1.300  & 1.447  &
  1.368 & 1.370 & 1.332   \\
$a_{2}$      &  $2^{+}$ & $^{3}P_{2}$ &  1.318$^{*}$   &
  1.320   & 1.320   & 1.319 & 1.317  \\
$\rho^{'}$   &  $1^{-}$ & $^{3}D_{1}$ & 1.47 &  1.482  &
  1.316  & 1.512 & 1.360  \\
?            &  $2^{-}$ & $^{3}D_{2}$ &         & 1.597   &
  1.504   & 1.622 & 1.532               \\
$\pi_{2}$    &  $2^{-}$ & $^{1}D_{2}$ & 1.670 & 1.607 &
  1.566 & 1.627  & 1.573    \\
$\rho^{''}$  &  $1^{-}$ & $^{3}S_{1}$ & 1.70 &  1.556  &
  1.451  & 1.570 & 1.460  \\
$\pi^{''}$   &  $0^{-}$ & $^{1}S_{0}$ &  1.77   & 2.042  &
  1.940 & 2.006 & 1.928   \\
\end{tabular}
\label{table6chap4}
\end{table}
\mediumtext
\begin{table}
\centering
\caption{Strange quarkonia masses in GeV for Breit and Salpeter models
         for the parameters as in Table~\protect\ref{table5chap4}.
         The calculated states are aligned with the observed states
         according to their spin parity. An asterisk on an observed value
         indicates a state employed in fitting.}
\smallskip
\begin{tabular}{ccc@{}dcccc}
  Meson   &  ${J}^{\pi}$ & $^{{2S+1}}L_{J}$  &  $~~~M_{expt}$ &
 $\stackrel{\textstyle {\rm Breit}}{\rm Mixed}$ &
 $\stackrel{\textstyle {\rm Breit}}{\rm Timelike}$ &
 $\stackrel{\textstyle {\rm Salpeter}}{\rm Mixed }$ &
 $\stackrel{\textstyle {\rm Salpeter}}{\rm Timelike}$
\\ \hline
?            &  $0^{-}$ & $^{1}S_{0}$ &    &  0.928  &
  0.929 & 0.640 &  0.741                       \\
$f_{0}$      &  $0^{+}$ & $^{3}P_{0}$ &  0.974  &
  1.401  &  1.198  & 1.422 & 1.164  \\
$\phi$       &  $1^{-}$ & $^{3}S_{1}$ &  1.019$^{*}$  &
  1.019  &  1.020  & 1.019 &  1.020    \\
$f_{0}$      &  $0^{+}$ & $^{3}P_{0}$ &  1.400  &  1.764  &
  1.570  & 1.784 & 1.859                 \\
$f_{1}$      &  $1^{+}$ & $^{3}P_{1}$ & 1.426   &  1.487  &
  1.346  & 1.501 &  1.369           \\
$f_{1}$      &  $1^{+}$ & $^{1}P_{1}$ &         &  1.515  &
  1.431  & 1.483 &  1.412           \\
$f_{2}^{'}$  &  $2^{+}$ & $^{3}P_{2}$ & 1.525   & 1.577   &
  1.587   & 1.581 &  1.536           \\
$f_{0}$      &  $0^{+}$ & $^{3}P_{0}$ &  1.587  &  2.647  &
  2.398  & 2.636 & 2.390 \\
?            &  $2^{-}$ & $^{3}D_{2}$ &         & 1.920   &
  1.777   & 1.927 &  1.788           \\
?            &  $2^{-}$ & $^{1}D_{2}$ &         & 1.928   &
  1.822   & 1.923 &  1.817           \\
\end{tabular}
\label{table7chap4}
\end{table}
%

\begin{references}
\bibitem{shifman79} M.~A. Shifman, A.~I. Vainshtein, and V.~I. Zakharov,
                    Nucl. Phys. {\bf B147}, 385 (1979).

\bibitem{reinders85}L.~J. Reinders, H. Rubinstein, and S. Yazaki,
                    Phys. Rep. {\bf 127}, 1 (1985).

\bibitem{rothe92}   H.~Rothe, {\it Lattice Gauge Theories:
                    An Introduction}, World
                    Scientific Lecture Notes in Physics, Vol. {\bf 43}.
                    World Scientific, Singapore, 1992.

\bibitem{gellmann64}M.~Gell-Mann,
                    Phys. Rev. Lett. {\bf 8}, 214 (1964).

\bibitem{godfrey89} S. Godfrey,
                    Nuovo Cimento {\bf 102A}, 1 (1989).

\bibitem{salpeter51}E.E. Salpeter and H.A. Bethe, Phys. Rev. 84, 1232 (1951).

\bibitem{long84}    C. Long, Phys. Rev. D {\bf 30}, 1970 (1984).

\bibitem{resag93}   J. Resag and D. Sch\"utte,
                    University of Bonn preprint TK-93-19.

\bibitem{parra94}   J. Parramore and J. Piekarewicz,
                    Nucl.~Phys.~{\bf A585}, 705 (1995).

\bibitem{piek92}    J. Piekarewicz, AIP Conference Proceedings
                    No.{\bf 269}, 412 (1992);
                    Rev. Mex. Fis. {\bf 39}, 542 (1993).

\bibitem{thou60}    D.J. Thouless, Nucl. Phys. {\bf 21}, 225 (1960);
                    Nucl. Phys. {\bf 22}, 78 (1961).

\bibitem{salpeter52}E.E. Salpeter, Phys. Rev. {\bf 87}, 328 (1952).

\bibitem{ullah71}   N. Ullah and D.J. Rowe, Nucl. Phys. {\bf A163},
                    257 (1971).

\bibitem{chi70}     B.E. Chi, Nucl. Phys. {\bf A146}, 449 (1970).

\bibitem{spence86}  J. Spence and J. Vary, Phys. Rev. D {\bf 35},
                    2191 (1987).

\bibitem{maung93}   K.M. Maung, D.E. Kahuna, and J. W. Norbury,
                    Phys. Rev. D {\bf 47}, 1182 (1993).

\bibitem{herbst77}  Ira W. Herbst, Commun. Math. Phys. {\bf 53}, 285 (1977).

\bibitem{raynal94}  J.C. Raynal, S.M. Roy, V. Singh, A. Martin, and J. Stubbe,
                    Phys. Lett. B {\bf 320}, 105 (1994).

\bibitem{lucha94}   Wolfgang Lucha and Franz F. Sch\"oberl, University
		    of Wien preprint UWThPh-1994-23; Hep-ph/9406312.

\bibitem{sakurai73} J.J. Sakurai, {\it Advanced Quantum Mechanics},
                    Addison-Wesley, 1973.

\bibitem{durand83}  B. Durand and L. Durand, Phys. Rev. D {\bf 28}, 396 (1983).

\bibitem{friar84}   J.L. Friar and E.L. Tomusiak, Phys. Rev. C {\bf 29}, 1537
                    (1984).

\bibitem{eichten7880} E.~Eichten, K.~Gottfried, T.~Kinoshita, K.~D.~Lane, and
                    T.~M.~Yan, Phys. Rev. D {\bf 17}, 3090 (1978);
                    Phys. Rev. D {\bf 21}, 203 (1980).

\bibitem{gara89}    Alan Gara, Bernice Durand, Loyal Durand, and
                    L.J. Nickisch, Phys. Rev. D {\bf 40}, 843 (1989);
                    Alan Gara, Bernice Durand, and Loyal Durand,
                    Phys. Rev. D {\bf 42}, 1651 (1990).

\bibitem{lagae92}   J.-F. Lag\"{a}e, Phys. Rev. D {\bf 45}, 317 (1992).

\bibitem{spence93}  J. Spence and J. Vary, Phys. Rev. C {\bf 47},
                    1282 (1993).

\bibitem{minuit92}  {\it MINUIT: Function Minimization and Error Analysis,
                    Version 92.1 (March 1992)},
                    CERN Program Library Long Writeup D506.
                    Application Software Group, Computing and Networks
                    Division, CERN, Geneva, Switzerland.

\bibitem{hikasa92}  K. Hikasa {\it et~al.}, Phys. Rev D {\bf 45}, S1 (1992).

\bibitem{armstrong92} E760 Collaboration, T.~A. Armstrong {\it et~al.},
                      Phys. Lett. {\bf 69}, 2337 (1992).

\bibitem{deboor78}  C. DeBoor, {\it A Practical Guide to Splines},
                    Springer, Berlin, 1978.

\bibitem{thooft76}  G. t'Hooft, Phys. Rev. D {\bf 14}, 34432 (1976).

\bibitem{shifman80} M.~A. Shifman, A.~I. Vainshtein, and V.~I. Zakharov,
                    Nucl. Phys. {\bf B163}, 46 (1980).

\bibitem{munz93}    J. Resag, C.R. M\"unz, B.C. Metsch, and H.R. Petry,
                    University of Bonn preprint TK-93-13;
                    C.R. M\"unz, J. Resag, B.C. Metsch, and H.R. Petry,
                    {\it ibid.,} TK-93-14.

\bibitem{gromes91}  W. Lucha, F.F. Sch\"oberl, and D. Gromes,
                    Phys. Rep. {\bf 200}, 127 (1991).

\bibitem{gross94}   F. Gross and J. Milana,
                    Phys. Rev. D {\bf 50}, 3332 (1994).

\bibitem{weinstein90} John Weinstein and Nathan Isgur, Phys. Rev. D {\bf 41},
                    2236 (1990).

\bibitem{arfken85}  G. Arfken, {\it Mathematical Methods in Physics}, 3rd ed.,
                    1985.

\bibitem{gradshteyn80} I.S. Gradshteyn and I.M. Ryzhik,
                    {\it Table of Integrals,
                    Series, and Products, corrected and enlarged edition},
                    Academic Press, New York (1980).
\end{references}
\end{document}